# Autonomous Minibus Service with Semi-on-demand Routes in Grid Networks

**Max T.M. Ng**
Graduate Research Assistant
Transportation Center
Northwestern University
600 Foster Street
Evanston, IL 60208, USA
Email: maxng@u.northwestern.edu

**Hani S. Mahmassani**
William A. Patterson Distinguished Chair in Transportation
Director, Transportation Center
Northwestern University
600 Foster Street
Evanston, IL 60208, USA
Email: masmah@northwestern.edu
Tel: 847.491.2276

## ABSTRACT

This paper investigates the potential of autonomous minibuses which take on-demand directional routes for pick-up and drop-off in a grid network of wider area with low density, followed by fixed routes in areas with demand. Mathematical formulation for generalized costs demonstrates its benefits, with indicators proposed to select existing bus routes for conversion with the options of zonal express and parallel routes. Simulations on modeled scenarios and case studies with bus routes in Chicago show reductions in both passenger costs and generalized costs over existing fixed-route bus service between suburban areas and CBD.

# INTRODUCTION

The flexibility and efficiency of demand responsive transit (DRT) in areas with sparse demand has been limited by the difficulty of human drivers to respond to real-time route changes and information. This required pre-booking trips and constrained routing optimization. Since the advent of smartphones and prevalence of on-demand transport services, the first limitation can be lifted by centralized online systems to accept ad-hoc service requests, manage routes, and communicate estimated arrival and journey times. While on-demand transit does not require autonomous functionality, we expect autonomous vehicles (AV) will bring more significant breakthroughs and provide a modification of this type of service. Centrally coordinated, autonomous buses can be routed real-time to requested points, resulting in an optimal journey time with respect to traffic conditions.

Opportunities for improvement exist in fixed-schedule fixed-route directional bus services *e.g.,* between suburban areas and downtown, by replacing the suburb section with demand-responsive routes as illustrated in **Figure 1**. The minibus travels around the suburb for pick-up and drop-off, then continues through the existing route downtown. It does not have to make compulsory stops, but instead optimizes the path based on requested stops.

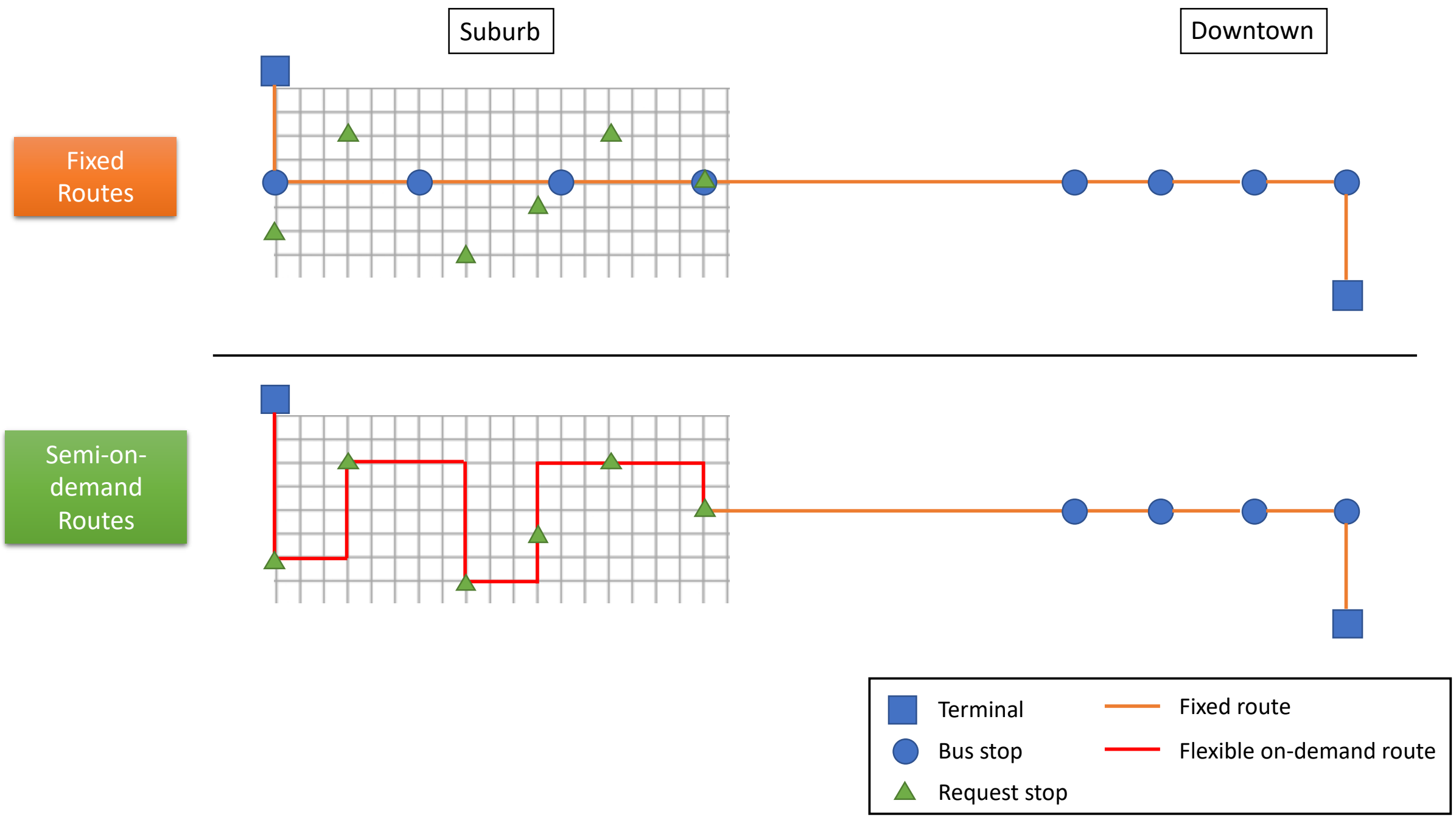

**Figure 1 - Illustration of Fixed and Semi-on-demand Routes**

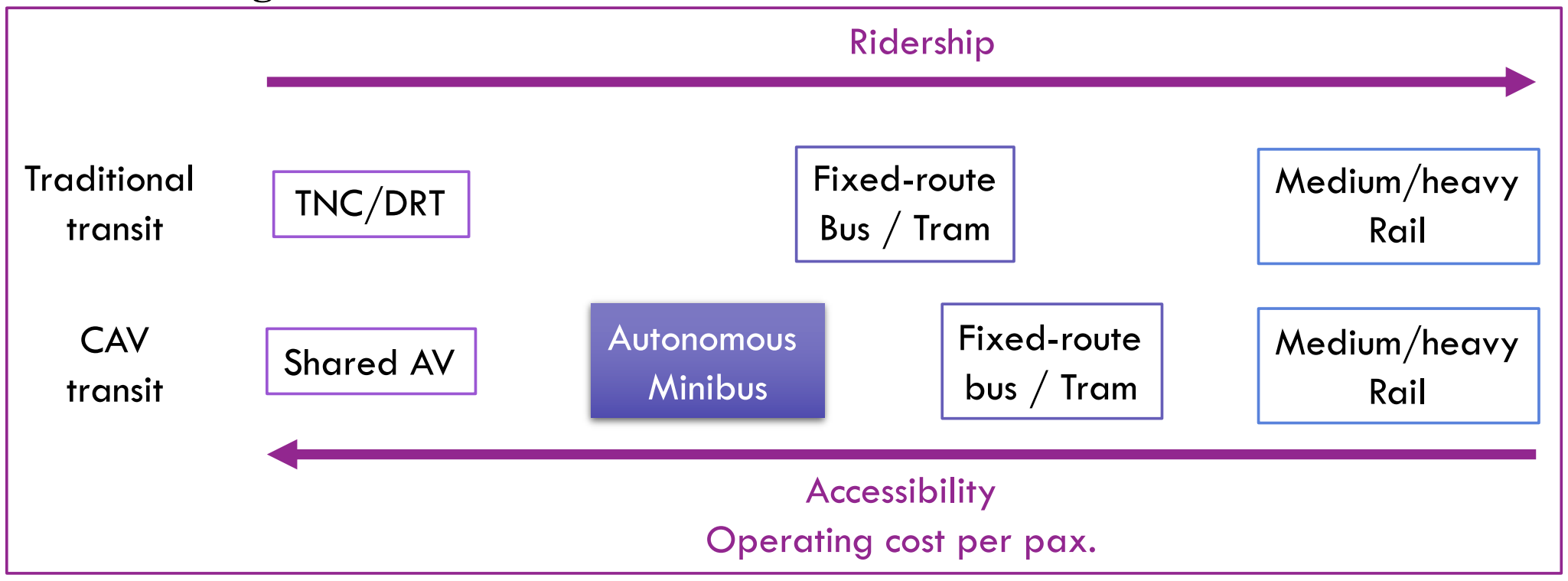

**Figure 2 - Positioning of the Proposed Autonomous Minibus Services**

The envisioned service could be competitive with fixed routes and shared autonomous vehicles (SAV) (**Figure 2**). Compared with fixed routes, it minimizes access distance, increases coverage area, and improves efficiency by reducing stop time for service stabilization. Besides, it also eliminates transfers for trunk-and-feeder networks, and associated fluctuations and scheduling difficulty. Over SAV with passenger cars, it provides a better economy of scale, particularly in routes with regular demand. It transports more passengers with fewer vehicles when fixed costs and downtown congestion are concerned.

This paper explores the benefits of this autonomous minibus service with semi-on-demand routes (AMSoD) by focusing on directional travel, *e.g.,* suburb-downtown, where constant demand justifies capacity higher than SAV but relatively sparse distribution of origin/destination discourages full-scale bus service. The generalized cost formulation, which is the sum of costs of access, waiting, riding, and operator, in a grid network is developed with the option to divide the service area for zonal express. To facilitate implementation, indicators are proposed to identify existing bus routes to adopt AMSoD based on demand and coverage with extension to parallel services. The findings are verified with simulation of general models and case studies with existing bus routes in Chicago.

This paper aims to address the following two questions:

1. Under what scenarios can the AMSoD service offer lower generalized costs and passengers' costs over current fixed routes?
2. Can the AMSoD service be expanded with zonal express or parallel service to cover more scenarios?

This section continues to introduce the service, followed by literature review in DRT, SAV, and bus networks. The paper proceeds with mathematical formulation of costs and selection indicators, which leads to the simulation model and case studies in Chicago. It then concludes with limitations and research directions.

**Service Description**

The proposed AMSoD service uses a central system to accept requests and coordinate buses. Passengers request service, informed with pick-up/drop-off and journey times. Instead of an ordinary fixed schedule at each bus stop, an average frequency (*e.g.,* 4/hour) is expected by the travelers. Meanwhile, the fixed routes downtown are maintained, considering limitations in bus stops, traffic regulations, and reduced edge of on-demand services due to the concentration of activities.

The service assumptions are as follows:

A1. The passengers' costs are lower than or the same as existing fixed-route services, so the demand patterns are the same.
A2. The number of minibuses assigned and departure headways are the same as those of fixed routes.
A3. All demands are within capacity.
A4. Travelers request services when needed and choose desirable pick-up/drop-off points or the closest possible points in case outside the service area.
A5. All passengers are ready for pick-up before bus arrival and picked up within a preset timeframe.
A6. A grid system as shown in **Figure 1** is assumed for the suburbs.
A7. Buses do not wait or travel backward more than one block for pick-up/drop-off.

While previous studies on semi-flexible service focused on maintaining service regularity and accessibility by keeping regular stops, technological advancement has promoted acceptance of on-demand services. The AMSoD service guarantees regularity with a semi-fixed schedule. Removing fixed stops further enhances service by reducing unnecessary bus segments. Stop request facilities can be provided in existing stops to bridge the digital divide.

Although the model in this paper does not explicitly require autonomous functionality, the service mode would be hardly feasible otherwise, for safety and reliability concerns to require human drivers to respond to real-time requests and immediate routing changes.

## LITERATURE REVIEW

### Demand Responsive Transit (DRT)

DRT problems were usually formulated as vehicle routing problems (VRP) with pick-up (many-to-one), delivery (one-to-many), and constraints of capacity and service level (*e.g.,* time window). It was known to be NP-hard similar to the traveling salesman problem. Errico et al. (*1*) proposed semi-flexible Demand Adaptive System which combined traditional fixed-route bus service with on-demand optional stops along the route to enhance flexibility while maintaining service regularity and accessibility without reservation. Various types of semi-flexible systems were identified (*2*).

Frei (*3*) pointed out the suitability of DRT at different times a day (not only focusing on peak hours) and in low-density suburbs, and described the flexible system design as a *"trade-off between operator and user generalized costs"* (p15). The proposed design process was to set up stops already recognized as origins and destinations with demand, then optimizing service. Performance measures were proposed to include in-vehicle travel time (IVTT), idle time, traveling time, time window violations, arrival/departure from compulsory stops, user delay, cost of random demand insertions, and vehicle utilization. The analysis recommended mandatory stops due to their small impact on ridership.

Cortés and Jayakrishnan (*4*) proposed a high-coverage point-to-point DRT which catered to multiple-origin-multiple-destination cases. The previous DRT was reviewed as less appealing due to the long waiting time for short trips and low average passenger load. The new model made use of heuristics to decide which vehicle to send and pick up passengers by considering metrics including level of service index (average passenger waiting time per average door-to-door IVTT), ride time index (average vehicle IVTT per average door-to-door IVTT), and average passenger load. The paper considered stochastic formulation but noted that strict real-time optimization would be computationally impossible.

Nourbakhsh & Ouyang (*5*) designed a new structured flexible-route transit system with pick-up and drop-off of each route through a pre-determined area, or a "tube" based on the framework of Daganzo (*6*). The collective network was then formed by individual flexible routes to serve the entire area. The operating performance was expressed in analytical functions of key design variables. The agency costs were estimated from total vehicle distance traveled per hour of operation and fleet size, noting that capital infrastructure investment is minimal for flexible transit. As for user cost, a higher level of service was provided by eliminating walking time. Evaluated with constrained nonlinear optimization to minimize system costs, results showed that low-to-moderate passenger demand would incur lower costs compared to fixed-route systems and taxis.

More recent efforts focus on integrating DRT with traditional fixed services. A dynamic switching policy between fixed-route and on-demand mobility was explored (*7*). Chen & Nie (*8*) studied supplementary improvement to fixed-route bus services with DRT to serve passengers with excessive access distance from bus stops in grid and radial networks.

Calabrò et al. (*9*) highlighted the limitations of fixed-schedule services in areas with sparse demand, due to excessive operating costs to maintain service quality, particularly in suburbs, leading to geographical inequity. In contrast, DRT was not suitable to serve dense demand for the potential tortuous routes and high operating costs. Agent-based simulation and continuous approximation showed that the combination of both would improve the service quality at off-peak and in suburbs while maintaining reasonable generalized costs.

**Shared Autonomous Vehicle (SAV) Services related to Autonomous Buses/Minibuses**

Numerous studies covered SAV with point-to-point services and passenger-car fleets. Narayannan et al. (*10*) summarized a number of recent literature, suggesting continuation of the combination of reservation-based and dynamic operations and identifying the research gap in revamping public transit systems with AV technologies.

Some ride-sharing studies bore similarities with DRT by considering high-capacity services (*11*). Dynamic trip assignment was considered with real-time optimal routes generated based on demand and vehicle locations. The problem was identified as a balance between fleet size, capacity, waiting time, travel delay, and operational cost, and validated with New York City taxi dataset.

Pinto et al. (*12*) looked into a joint transit network redesign and SAV fleet size determination problem with bi-level mathematical programming. The upper-level problem solved for the fleet size and set the frequency with a nonlinear solver, while the lower-level problem identified mode choice with iterative agent-based simulation. An example of traveler's demand and existing multimodal network in Chicago showed improvement in waiting time without increasing operators' costs.

A completely autonomous transit system was proposed by Winter et al. (*13*) to replace the existing transit system with demands of low-to-moderate levels and many-to-many patterns. Model simulation was carried out with cost assumptions of electric vehicles and sensitivity tests on capacities (2-40 passengers), demand level, vehicle dwell time, and initial vehicle locations for assessing impacts on operational costs and passenger generalized costs. The operating cost was found to be around EUR 0.84-1.22/passenger-trip. Besides, the optimal vehicle capacity was four passengers, which was probably related to the distributed origin and destination patterns.

Autonomous bus was an evolving topic with on-going research efforts. Despite autonomous shuttles in trial or operation, various forms of autonomous bus services were envisioned to improve existing transit.

The optimization of mixed-fleet ordinary and autonomous buses was explored with particular features including flexibility to assemble or disassemble autonomous minibuses and adjustment of headway by controlling the running time (*14*). The problem was formulated as nonlinear integer programming and solved with a model of dynamic programming. Dynamic stochastic experiments were carried out regarding dispatching policy, autonomous bus penetration rate, bus running time variation, and demand level. An increasing proportion of autonomous buses in the fleet benefitted both operators and passengers.

Cao and Ceder (*15*) proposed autonomous shuttle bus service with methodology to optimize timetabling and vehicle scheduling, considering potential stop skipping and the graphical concept of deficit function as the number of additional vehicles required. The model found savings in total passenger travel time and the number of vehicles over existing routes. Similarly, ordinary fixed routes with skipping capabilities were proposed based on origin-destination generated from payment data in Zhejiang, China (*16*). Agent-based simulation, which incorporated the interaction of passengers, buses (in-operation/idle), and control center, concluded higher utilization of buses and better adaptiveness to change in demand.

There has also been recent research on the behavioral side of rolling out autonomous bus service. A qualitative survey in Finland pointed out that the most important reasons for the decision to use an autonomous bus are route and flexibility (*17*). On-demand service will be made possible with AVs by ruling out human drivers confused by route alternation. Basu et al. (*18*) suggested the necessity of mass transit to address mobility needs despite the development of automated mobility-on-demand and echoed the importance of flexible routes. By eliminating driver costs, Bösch et al. (*19*) suggested future directions for autonomous buses with lower capacities and higher frequencies. Cost analysis of predicted autonomous mobility services resulted in a threshold of 21 passengers between a minibus and a bus. As for acceptability of this new technologies in public transit, Pigeon et al. (*20*) concluded that factors included service characteristics such as schedules and times. Specifically for minibuses, Bernhard et al. (*21*) found out general positive evaluation by the respondents and the role of experience on its acceptance.

**Summary and Related Bus Network Research**

Previous literature on autonomous buses primarily focused on fully flexible routes or supplementary DRT to existing fixed-route networks. A research gap for the extent of benefits and applications for medium/high-capacity SAV services, *i.e.* AMSoD, in public transit systems was noted. The proposed service combines existing DRT research and bus services. The benefits brought by SAV services can overcome the existing downsides, in particular in low-to-moderate demand areas such as downtown-suburb linear corridors.

For this kind of linear corridor with flows converging into a main avenue, Gschwender et al. (*22*) compared the performance of feeder-trunk, direct, exclusive, and shared service structures. Transfer penalty was found to be the key to deciding on dominance of the feeder-trunk network. Furthermore, with more long trips, other network types were more attractive for reduced transfer, idle capacity, and waiting time.

Zhang et al. (*23*) compared on-demand public buses set up based on pre-existing demands with limited stops with park-and-ride services on a CBD-suburb corridor in terms of consumer surplus and operating net profit. The on-demand service was more favorable for long-distance commuting trips.

The problem characteristics and solution methods of various studies in the literature are summarized in **Table 1**.

**Table 1 - Problem Characteristics and Solution Method in Literature**

| Ref. | Nature | Vehicle type | Route | Trip pattern | Solution Method |
|---|---|---|---|---|---|
| (Alonso-Mora et al. 2017) | Ridesharing | Heterogeneous up to 10 pax | Flexible | Multiple-origin-multiple-destination | Greedy assignment with constrained optimization |
| (Calabrò et al. 2021) | "Flexible Transit' with a switch between fixed-schedule and DRT | Bus | Switch between fixed and DR | Feeder | Continuous Approximation |
| (Cao and (Avi) Ceder 2019) | Autonomous bus | Bus sample capacity of 60 | Fixed (with stop skipping) | Original bus route | Binary variable iteration with genetic algorithm |
| (Chen and Nie 2017) | DRT | Bus | DR line connecting to fixed routes stop | Feeder | Mixed integer programming |
| (Cortés and Jayakrishnan 2002) | DRT | 7-seat vans or lower | Flexible | Multiple-origin-multiple-destination | Simulation with heuristics |
| (Dai et al. 2020) | Autonomous DRT | Bus | Fixed | Original bus route | Integer nonlinear programming with dynamic programming |
| (Edwards and Watkins 2013) (*24*) | DRT | Bus | Flexible in a griddled street system | Multiple-origin-multiple-destination | Simulation |
| (Errico et al. 2021) | DRT | Bus | Semi-flexible (compulsory + optional stops) | Multiple-origin-multiple-destination | Simulation |
| (Frei 2015) | DRT | Bus | Semi-flexible (compulsory + optional stops) | Feeder | Simulated annealing heuristic |
| (Nourbakhsh and Ouyang 2012) | DRT | Bus | Flexible in a tube | Multiple-origin-multiple-destination | Constrained nonlinear optimization |
| (Pinto et al. 2020) | SAMS with bus and rail | SAMS | Flexible | Multiple-origin-multiple-destination | Bi-level programming with non-linear solver and simulation |
| (Tong et al. 2017) (*25*) | DRT | Bus capacity of 10 | Flexible | Multiple-origin-multiple-destination | Lagrangian decomposition with a space-time prism-based method |
| (Winter et al. 2018) | Autonomous DRT | Vehicle capacity of 2-40 | Flexible | Multiple-origin-multiple-destination | Simulation |
| (Zhai et al. 2020) | Autonomous bus | Bus | Fixed | Multiple-origin-multiple-destination | Agent-based simulation |
| (Zhang, Wang, and Meng 2018) | DRT | Bus | Fixed on a linear travel corridor | Suburb-downtown | Heuristic |
| This paper | Autonomous Bus/DRT | Minibus capacity of 15-30 | Semi-on-demand directional | Suburb-downtown | Simulation |

## MATHEMATICAL FORMULATION

As discussed in Introduction, the network in the suburb section is modeled as a grid-like structure in **Figure 3** similar to (*5*). With the assumption of a maximum access time, the locations of demands are spatially distributed around an existing bus stop. Then, service coverage areas are delineated in the grids as shown. Meanwhile, the temporal distribution of demand is random.

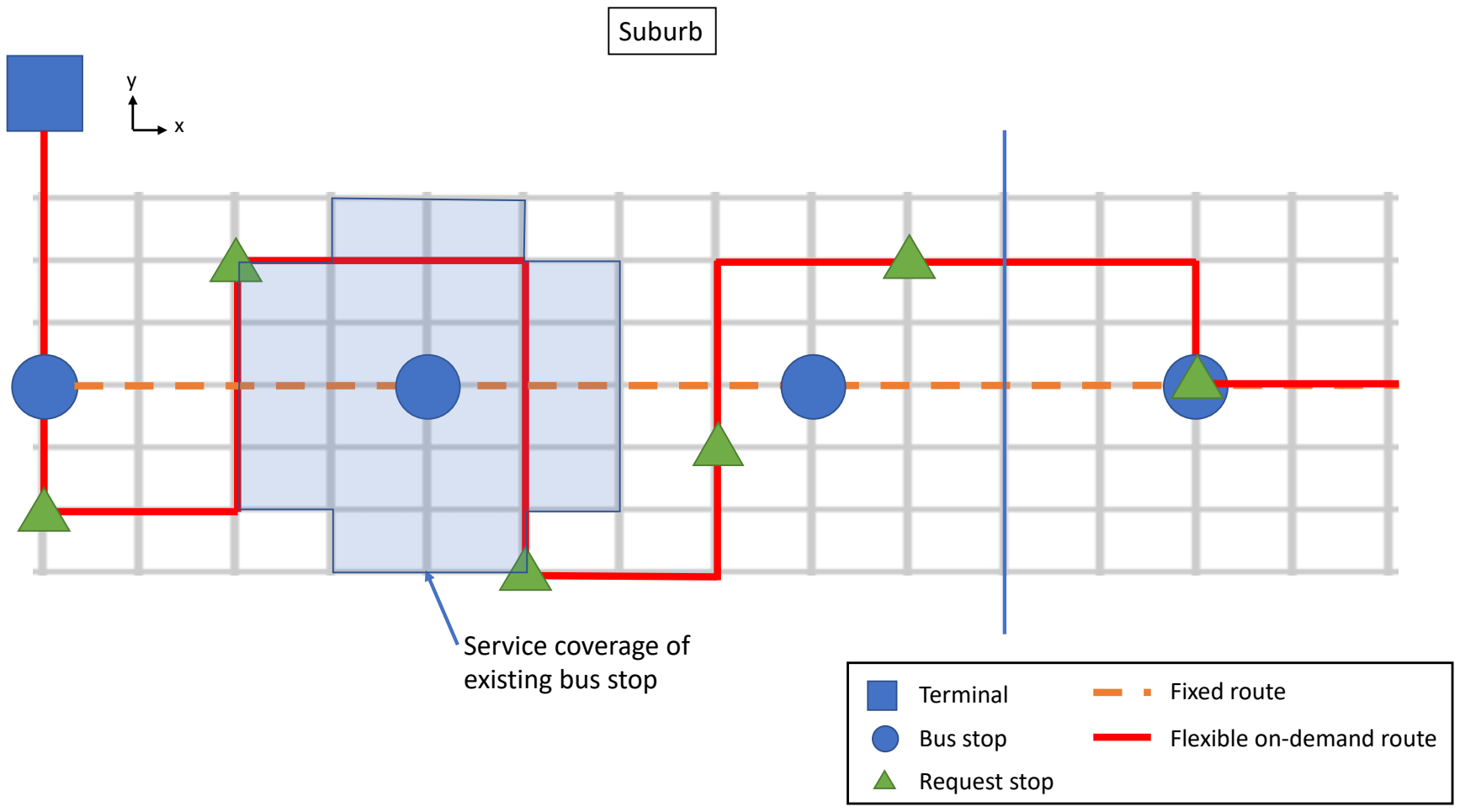

**Figure 3 - Grid Model of Suburbs**

The perpendicular displacement of semi-on-demand routes is first discussed, followed by general cost formulations for both services and cost comparison. Parameters are summarized in **Table 2**. The x-direction is defined as bus traveling direction, with y-direction as the perpendicular.

**Table 2 - Notations of Model Parameters**

| Category | Notation | Description |
|---|---|---|
| Cost | $\gamma_w$ | Penalty multiplier of waiting time |
| | $\gamma_a$ | Penalty multiplier of access time |
| | $\gamma_r$ | Penalty multiplier of riding time; set as 1 |
| | $\gamma_o$ | Operator's unit cost of vehicle distance |
| | $VOT$ | Value of time |
| | $s_{jk}$ | Access time of passenger *k* for fixed-route bus service |
| | $C_{w,j}$ | Waiting time cost of trip *j* of fixed-route bus service |
| | $C'_{w,j}$ | Waiting time cost of trip *j* of AMSoD service |
| | $C_{a,j}$ | Access time cost of trip *j* of fixed-route bus service |
| | $C'_{a,j}$ | Access time cost of trip *j* of AMSoD service |
| | $C_{r,j}$ | Riding time cost of trip *j* of fixed-route bus service |
| | $C'_{r,j}$ | Riding time cost of trip *j* of AMSoD service |
| | $C_{o,j}$ | Operator cost of trip *j* of fixed-route bus service |
| | $C'_{o,j}$ | Operator cost of trip *j* of AMSoD service |
| | $TC$ | Hourly generalized cost of fixed-route bus service |
| | $TC'$ | Hourly generalized cost of AMSoD service |
| | $\Delta TC$ | Hourly generalized cost of AMSoD service minus that of fixed-route bus service |

| Category | Notation | Description |
|---|---|---|
| | $\Delta TC_j$ | Generalized cost of bus trip *j* of AMSoD service minus that of fixed-route bus service |
| | $TC_z{}'(n)$ | Hourly generalized cost of AMSoD service in zonal express service with *n* zones |
| | $TC_p{}'$ | Hourly generalized cost of AMSoD service in parallel routes |
| | $\Delta TC_p$ | Hourly generalized cost of AMSoD service in parallel routes minus that of fixed-route bus service |
| Distance | $d'_{x,j}$ | x-direction portion of journey distance of trip *j* |
| | $d'_{y,j}$ | y-direction portion of journey distance of trip *j* of AMSoD service |
| | $x_{jk}$ | x-direction demand position of passenger *k* |
| | $y_{jk}$ | y-direction demand position of passenger *k* |
| | $L_x$ | x-direction length of a grid |
| | $L_y$ | y-direction length of a grid |
| | $GL_x$ | x-direction suburb length |
| | $GL_y$ | Half of the y-direction coverage area of bus service |
| | $d_{x,jk}$ | Parallel portion of journey distance traveled by passenger *k* of trip *j* in fixed-route bus service |
| | $d'_{x,jk}$ | Parallel portion of journey distance traveled by passenger *k* of trip *j* in AMSoD service |
| | $d'_{y,jk}$ | Perpendicular portion of journey distance traveled by passenger *k* of trip *j* in AMSoD service |
| | $MD(Y)$ | Mean absolute difference (MD) of y-directional positions of demand |
| Time | $IVTT_j$ | In-vehicle Travel Time of trip *j* of fixed-route bus service |
| | $IVTT'_j$ | In-vehicle Travel Time of trip *j* of AMSoD service |
| | $IVTT_{jk}$ | In-vehicle Travel Time of passenger *k* on trip *j* of fixed-route bus service |
| | $IVTT'_{jk}$ | In-vehicle Travel Time of passenger *k* on trip *j* of AMSoD service |
| | $s_{ak}$ | Access time of passenger *k* for fixed-route bus service |
| | $s_{wk}$ | Waiting time of passenger *k* for fixed-route bus service |
| | $s'_{wk}$ | Waiting time of passenger *k* for AMSoD service |
| | $s_o$ | Maximum access time for passengers to take the bus |
| | $t_s$ | Dwell time per stop of fixed-route bus service |
| | $t'_s$ | Dwell time per stop of AMSoD service |
| | $t_k$ | Passenger *k*'s time to start the trip (walk to the bus stop for fixed-route bus services, or ready for picking up for AMSoD service) |
| | $t_{b,s}$ | Arrival time of the bus at stop *s* of fixed-route bus service |
| | $t'_{ak}$ | Actual pick-up time of passenger *k* of AMSoD service |
| Speed | $v_w$ | Average walking speed of travelers |
| | $v_d$ | Average speed of buses in suburb section |
| | $v_H$ | Average speed of buses on highway (for zonal express) |
| Passenger | $K_j$ | Number of passengers in trip *j* |
| | $\lambda$ | Number of passengers per hour |

| Category | Notation | Description |
|---|---|---|
| | $N_k$ | Number of stops traveled by passenger *k* in the suburb of fixed-route bus service |
| Bus | $NS$ | Number of stops in the suburb of fixed-route bus service |
| | $d_{xs}$ | Station spacing in x-direction |
| | $bs_k$ | Bus stop for passenger *k* in fixed-route bus service |
| | $x_{bs,s}$ | x-direction position of bus stop *s* in fixed-route bus service |
| | $H$ | Headway / departure headway for the AMSoD service (combined headway for zonal express and parallel routes) |
| | $n$ | Number of zones in zonal express services |
| | $n_o$ | Optimal number of zones for zonal express services |

**Perpendicular Distance of Semi-on-demand Routes**

**Figure 4** shows an example of y-directional displacement between demand. As per Service Assumption A7, the minibus service travels mainly along the x-direction and detour in the y-direction to reach request points. Therefore, the x-directional portion of travel distance is the same as that of fixed routes, while the y-directional portion equals the displacement between request points (Eq. (1)).

Eq. (2) evaluates the expected y-directional distance based on the mean absolute difference (MD) between stops assuming symmetric spatial distribution. Eq. (3) estimates MD under uniform and normal distribution of demand. As demand at extremes from existing bus lines greatly increases perpendicular distance, demand points distributed normally giving lower MD result in shorter y-directional travel distance over uniform distribution. The general formula is extended for individual travelers in Eq. (4).

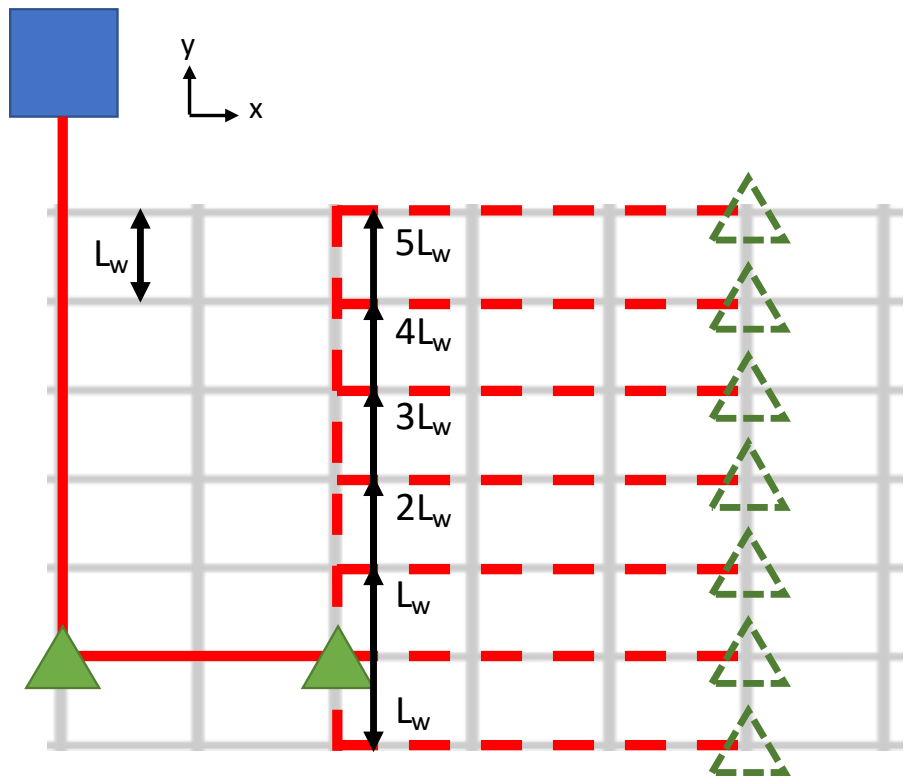

**Figure 4 - Illustration of Perpendicular Portion of Journey Distance**

$$d'_{y,j} = |y_{j,1}| + \sum_{k=2}^{K_j} |y_{j,k} - y_{j,k-1}| + |y_{j,K_j}| \tag{1}$$

$$E[d'_{y,j}] = |y_{j,1}| + |y_{j,K_j}| + (K_j - 1)MD(Y) \approx K_j MD(Y) \tag{2}$$

$$MD(Y) = E[|Y_i - Y_{i-1}|] = \begin{cases} \frac{1}{3}(b-a) \; for \; Y \sim U(a,b) \\ \frac{2}{\sqrt{\pi}} \sigma \; for \; Y \sim N(0,\sigma) \end{cases} \tag{3}$$

$$d'_{y,jk} = \sum_{m=k+1}^{K_j} |y_{j,m} - y_{j,m-1}| + |y_{j,K_j}| \approx (K_j - k) MD(Y) \tag{4}$$

**Costs**

The cost formulation of fixed route and semi-on-demand generally follows Newell's (*26*), where the generalized costs, $TC$, are the sum of passenger's costs (access, waiting, and riding) and operator's cost (Eq. (5)).

The focus is on the difference in costs between the two services, so only the costs in the suburb section are considered ignoring the identical section beyond suburbs. To ease comparison across cost types, penalty multipliers are introduced to access and waiting costs, while setting that of riding cost as unity.

The costs of pick-up trips originated from suburbs refer unless otherwise noted, but the costs of drop-off trips are similar except the variance components of waiting time shall be smaller in semi-on-demand cases.

*Traditional Fixed Routes*

The costs of pick-up and drop-off trips for fixed routes are both the sum of access, waiting, riding, and operator's costs (Eq. (5)). The access cost is estimated from access time proportional to access distance between demand origin to the bus stop (Eq. (6) and (7)). The maximum access time ($s_o$) bounds the access distance and defines the coverage area. The waiting cost is evaluated by the waiting time experienced by passengers, *i.e.,* time difference between passenger and bus arrival at stops (Eq. (8) and (9)). The riding cost is derived from IVTT as journey distance divided by bus speed plus dwell time at stops (Eq. (10) and (11)). The operator's costs are estimated in a per vehicle-distance basis (Eq. (13)).

$$TC_j = C_{a,j} + C_{w,j} + C_{r,j} + C_{o,j} \tag{5}$$

$$C_{a,j} = \gamma_a \, VOT \sum_k s_{ak} \approx \gamma_a VOT \, K_j \overline{s_{ak}} \tag{6}$$

$$s_{ak} = \frac{|x_{bs,bs_k} - x_{j,k}| + |y_{j,k}|}{v_w} \leq s_o = \frac{GL_y}{v_w} \tag{7}$$

$$C_{w,j} = \gamma_w VOT \sum_k s_{wk} \tag{8}$$

$$s_{wk} = t_k + s_{ak} - t_{b,bs_k} \tag{9}$$

$$C_{r,j} = \gamma_r \, VOT \sum_k IVTT_{jk} \tag{10}$$

$$IVTT_{jk} = \frac{d_{x,jk}}{v_d} + t_s \, (N_k - bs_k - 1) \tag{11}$$

$$d_{x,jk} = GL_x - x_{bs,bs_k} \tag{12}$$

$$C_{o,j} = \gamma_o d_{x,j} = \gamma_o L_x \tag{13}$$

*Semi-on-demand Routes*

The costs of semi-on-demand routes differ from fixed routes mainly by the absence of access cost (Eq. (15)) and y-directional detour. The waiting time is the difference between journey start time and actual pick-up time (Eq. (17)). The IVTT can be separated into the x-directional journey time of fixed routes and y-directional detour to connect the pick-up/drop-off locations, as well as the dwell time for every passenger pick-up point (Eq. (19)). This also applies to the operator's cost (Eq. (20)).

$$TC'_j = C'_{a,j} + C'_{w,j} + C'_{r,j} + C'_{o,j} \tag{14}$$

$$C'_{a,j} = 0 \tag{15}$$

$$C'_{w,j} = \gamma_w VOT \sum_k s'_{wk} \tag{16}$$

$$s'_{wk} = t'_{ak} - t_k \tag{17}$$

$$C'_{r,j} = \gamma_r \, VOT \sum_k IVTT'_{jk} \tag{18}$$

$$IVTT'_{jk} = \frac{\left(d'_{x,jk} + d'_{y,jk}\right)}{v_d} + t'_s(K_j - k - 1) \tag{19}$$

$$C'_{o,j} = \gamma_o\left(d_{x,j} + d'_{y,j}\right) \tag{20}$$

*Expected Waiting Time, Riding Time, and Total Costs*

With the previously derived cost components, we can further estimate the expected values of waiting and riding time for both fixed and semi-on-demand routes by assuming a symmetric spatial distribution of demand. The expected waiting time of a passenger is the sum of half expected headway and headway variance per half expected headway (Eq. (21)). A further assumption of Poisson arrivals simplifies the expectation term to design headway. The expected IVTT is taken as the travel distance at the mid-point of service area, which consists of average travel times in y-direction and x-direction as well as stopping time (Eq. (22) and (23)). The mean y-directional travel distance is equal to half of the product of the number of passengers and MD in Eq. (3). Summing all components of costs results in the total costs in Eq. (24) and (25), and allow derivation of a more tractable expression of total hourly generalized costs in Eq. (26) and (27) under division by headway.

$$E[s_{wk}] = E[s'_{wk}] = \frac{E[H]}{2} + \frac{\sigma_{Hk}^2}{2E[H]} = \frac{H}{2} + \frac{\sigma_{Hk}^2}{2H} \tag{21}$$

$$E\left[IVTT_{jk}\right] \approx \frac{L_x}{2v_d} + \frac{t_s\ NS}{2} \tag{22}$$

$$E\left[IVTT'_{jk}\right] \approx \frac{L_x}{2v_d} + \frac{K_j}{2v_d}MD(Y) + \frac{t'_s K_j}{2} \tag{23}$$

$$\mathrm{TC_j} \approx \gamma_a VOT\ K_j\overline{s_{ak}} + \gamma_w VOT\ K_j\left(\frac{H}{2} + \frac{\sigma_{Hk}^2}{2H}\right) + \gamma_r VOT\ K_j\left(\frac{L_x}{2v_d} + \frac{t_s\ NS}{2}\right) + \gamma_o L_x \tag{24}$$

$$\begin{aligned}\mathrm{TC'_j} \approx\ & \gamma_w VOT\ K_j\left(\frac{H}{2} + \frac{\sigma_{Hk}^2}{2H}\right) + \gamma_r VOT\ K_j\left(\frac{L_x}{2v_d} + \frac{K_j}{2v_d}MD(Y) + \frac{t'_s K_j}{2}\right) \\ & + \gamma_o\left(L_x + K_j MD(Y)\right)\end{aligned} \tag{25}$$

$$\mathrm{TC} \approx \gamma_a VOT\ \lambda\overline{s_{ak}} + \gamma_w VOT\ \lambda\left(\frac{H}{2} + \frac{\sigma_{Hk}^2}{2H}\right) + \gamma_r VOT\ \lambda\left(\frac{L_x}{2v_d} + \frac{t_s\ NS}{2}\right) + \gamma_o\left(\frac{L_x}{H}\right) \tag{26}$$

$$\begin{aligned}\mathrm{TC'} \approx\ & \gamma_w VOT\ \lambda\left(\frac{H}{2} + \frac{\sigma_{Hk}^2}{2H}\right) + \gamma_r VOT\ \lambda\left(\frac{L_x}{2v_d} + \frac{\lambda H}{2v_d}MD(Y) + \frac{t'_s\lambda H}{2}\right) \\ & + \gamma_o\left(\frac{L_x}{H} + \lambda\ MD(Y)\right)\end{aligned} \tag{27}$$

*Zonal Express for Semi-on-demand Routes*

Eq. (23) and (27) suggest that high demand and y-directional distance increase the IVTT and generalized costs of semi-on-demand routes. An advantage of minibuses is to capitalize on the low capacities to divide the coverage area. Minibuses in the first zone skip other stops before heading downtown (**Figure 5**) similar to traditional zonal express.

By the assumptions of symmetric demand spatial distribution, insignificant additional waiting time incurred by headway variance, and higher highway speed outside grids, Eq. (28) is the total hourly generalized costs of the whole service region. The waiting cost increases due to greater headway for each zone using the same total number of buses. The riding cost is smaller due to reduced y-directional detour and faster traveling on highway. The distance-based operator cost is lower by saving x-directional travel of some buses.

Then by the partial derivative of $\mathrm{TC}'_z(n)$ (Eq. (29)) and ignoring the uncertainty of headway, the number of zones to minimize the total cost is found in Eq. (30). The optimal $n_o$ increases with smaller original headway, longer x-directional distance, higher speed of highway over local streets, and higher operator's cost.

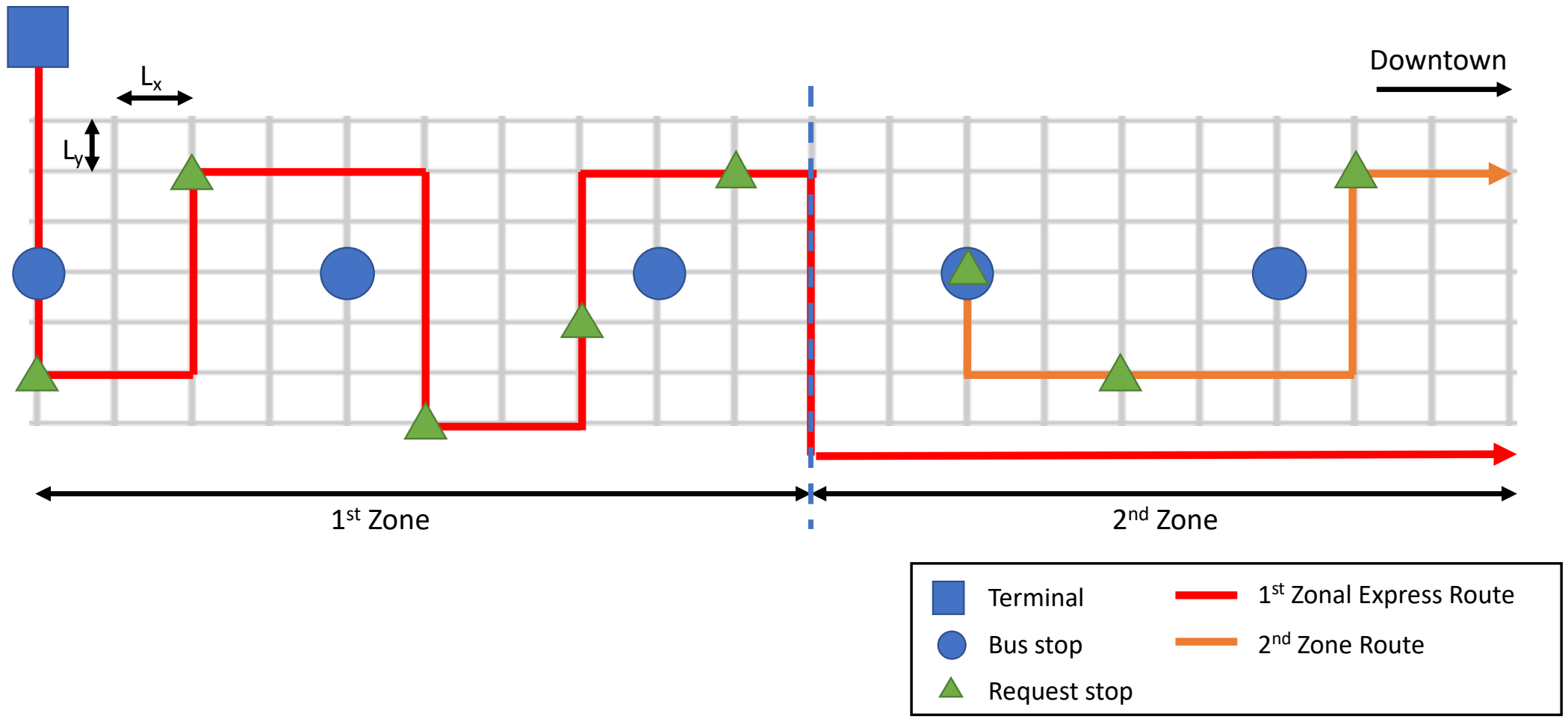

**Figure 5 - Illustration of Zonal Division**

$$\mathrm{TC}_z'(n) = \gamma_w VOT\, n\lambda\left(\frac{H}{2}+\frac{\sigma_{Hk}^2}{2H}\right)+\gamma_r VOT\,\lambda\left[\frac{L_x}{2nv_d}+\frac{L_x(n-1)}{2nv_H}+\lambda H\frac{MD(Y)}{2v_d}+\frac{t_s'\lambda H}{2}\right] + \gamma_o\left(\frac{L_x}{2H}\frac{n+1}{n}+\lambda\, MD(Y)\right) \quad (28)$$

$$\frac{\partial TC_z'(n)}{\partial n} \approx \gamma_w VOT\frac{\lambda H}{2}-\gamma_r VOT\frac{\lambda}{n^2}\left(\frac{L_x}{2v_d}-\frac{L_x}{2v_H}\right)-\gamma_o\frac{L_x}{2n^2H}=0 \quad (29)$$

$$n_o=\sqrt{\frac{L_x}{\gamma_w H}\left[\gamma_r\left(\frac{1}{v_d}-\frac{1}{v_H}\right)+\frac{\gamma_o}{\lambda H\, VOT}\right]} \quad (30)$$

## Replacement of Current Fixed Routes

To facilitate implementation of the AMSoD service, this section focuses on identifying scenarios where it could reduce generalized costs over fixed routes, with AV and same number of minibuses in both cases. The study area is assumed to be currently served with an off-peak direct bus route downtown for comparison. The service time of interest would be off-peak with relatively sparse demand. Competitive service can be extended to peak hours with a higher frequency or zonal express.

The key saving of AMSoD is access cost. However, while the x-directional travel distance and time are the same within grids, extra y-directional distance is incurred to reach request points. This leads to higher waiting, riding, and operator's costs. The difference in expected waiting time is caused by the headway variance induced by extra journey time before reaching the stop (Eq. (21)). The main components of the variance in IVTT are the additional y-directional distance traveled and stopping time for previous pick-ups (Eq. (31)), where the former is evaluated by the second central moment and the latter is derived based on a Poisson process of the number of passengers on board before $k$, *i.e.* $Poisson(K_j x_k/L_x)$ .

$$\sigma_{Hky}^2 = Var\left[IVTT_{jy}' - IVTT_{jky}'\right] \approx \frac{1}{K_j}\sum_{k=1}^{K_j}\left(\frac{k}{v_d}MD(Y) - \frac{K_j}{2v_d}MD(Y)\right) + \frac{K_j(L_x/2)}{L_x}{t_s'}^2$$
$$\approx \left(\frac{MD(Y)}{v_d}\right)^2\frac{K_j^2 + 6K_j + 2}{12} + \frac{K_j}{2}{t_s'}^2 \quad (31)$$

The difference in generalized costs can be evaluated in terms of the number of passengers and distribution of demand in Eq. (32) and (33), with an assumption of similar stopping times. $\Delta \mathrm{TC_j} < 0$ implies the saving in access cost is greater than other terms and the semi-on-demand route is beneficial. Between the trade-off of access and waiting, riding, and operator's cost, $\Delta TC$ is subject to the riding cost (from the detour to request points) quadratic to the number of passengers. Besides, higher perpendicular dispersion of demand ($MD(Y)$) is a negative factor which increases all three costs. Furthermore, vehicle speed is a beneficial factor for the proposed service. Sparser areas with less busy traffic would favor its performance.

Here we introduce two indicators. The Selection Indicator (SI) in Eq. (34) is the ratio of additional costs per unit of access cost saving. The condition for the proposed service to be favorable without detailed simulation is $SI_j < 1$. Similarly, Eq. (35) shows a simplified advantageous upper bound of the demand density ignoring the randomness portion of waiting cost. This notation indicates that the product of hourly demand and headway determines the attractiveness of the AMSoD service. While the service might attract induced demand due to flexibility and advantages brought to passengers, the extra demand would not render the service less attractive as long as the frequency can be increased accordingly. This improvement of service quality is justifiable by the increase in ridership.

$$\Delta \mathrm{TC_j} \approx VOT\left(-\gamma_a K_j \overline{s_{ak}} + \gamma_w \frac{K_j}{2H}\left(\left(\frac{MD(Y)}{v_d}\right)^2\frac{K_j^2 + 6K_j + 2}{12} + \frac{K_j}{2}{t_s'}^2\right) + \frac{\gamma_r K_j^2 MD(Y)}{2v_d}\right) + \gamma_o K_j MD(Y) \quad (32)$$

$$\Delta \mathrm{TC} \approx \frac{VOT}{H}\left(-\gamma_a \lambda H\,\overline{s_{ak}} + \gamma_w \frac{\lambda}{2}\left(\left(\frac{MD(Y)}{v_d}\right)^2\frac{(\lambda H)^2 + 6\lambda H + 2}{12} + \frac{\lambda H}{2}{t_s'}^2\right) + \frac{\gamma_r (\lambda H)^2 MD(Y)}{2v_d}\right) + \gamma_o \lambda\, MD(Y) \quad (33)$$

$$SI = \frac{\frac{\gamma_r \lambda H\, MD(Y)}{2v_d} + \frac{\gamma_w}{2H}\left(\left(\frac{MD(Y)}{v_d}\right)^2\frac{(\lambda H)^2 + 6\lambda H + 2}{12}\right) + \frac{\gamma_w \lambda}{4}{t_s'}^2 + \frac{\gamma_o}{VOT}MD(Y)}{\gamma_a \overline{s_{ak}}} \quad (34)$$

$$\lambda H < \frac{2\gamma_a \overline{s_{ak}} v_d}{\gamma_r MD(Y)} - \frac{2\gamma_o v_d}{\gamma_r VOT} \quad (35)$$

*Parallel Routes*

For existing fixed routes with wider access coverage (*i.e.,* high $MD(Y)$), the introduced single semi-on-demand route may not be favorable due to excessive y-directional detour. This can be solved by dividing it into two parallel semi-on-demand routes, each of which covers the demand points in a parallel stretch of area (**Figure 6**).

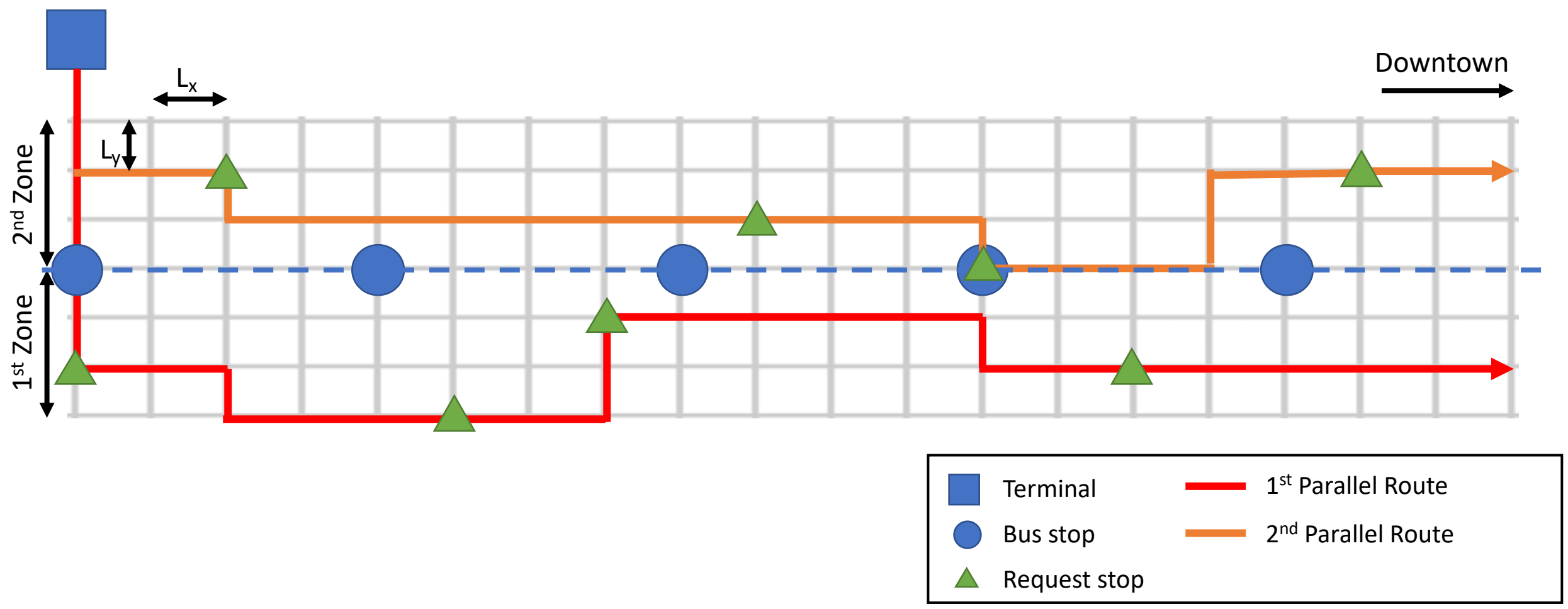

**Figure 6 - Illustration of Parallel Routes**

The generalized cost change between a fixed route and two parallel semi-on-demand routes shown in Eq. (36) is evaluated based on double headway for passengers and roughly half y-directional MD. This increases the deterministic components of waiting cost and reduces its stochastic components, as well as riding and operator's costs. Eq. (37) and (38) are updated SI and upper bound of demand, which is higher than the previous bound in Eq. (35) for high $MD(Y)$ cases.

$$\Delta \mathrm{TC}_p \approx \frac{VOT}{H}\left(-\gamma_a \lambda H \overline{s_{ak}} + \gamma_w \frac{\lambda H^2}{2} + \gamma_w \frac{\lambda}{2}\left(\left(\frac{MD(Y)}{2v_d}\right)^2 \frac{(\lambda H)^2 + 6\lambda H + 2}{12} + \frac{\lambda H}{2} {t'_s}^2\right) + \gamma_r \frac{(\lambda H)^2 MD(Y)}{4v_d}\right) + \gamma_o \lambda\, MD(Y) \tag{36}$$

$$SI_p = \frac{\gamma_w \left(\frac{H}{2} + \frac{1}{2H}\left(\left(\frac{MD(Y)}{2v_d}\right)^2 \frac{(\lambda H)^2 + 6\lambda H + 2}{12}\right) + \frac{\lambda}{4} {t'_s}^2\right) + \gamma_r \frac{\lambda H\, MD(Y)}{4v_d} + \gamma_o \frac{MD(Y)}{VOT}}{\gamma_a \overline{s_{ak}}} \tag{37}$$

$$\lambda H < \frac{2v_d}{\gamma_r MD(Y)}\left(2\gamma_a \overline{s_{ak}} - \gamma_w H\right) - \frac{4\gamma_0 v_d}{\gamma_r VOT} \tag{38}$$

For zonal express, the saving in travel time relies on reduction in stopping time and speed advantage of highway over local streets. For cases in grids of local streets and on-demand stops, the time saving is limited when compared to parallel routes or simply increasing frequency of the same route to meet the demand.

**Figure 7** summarizes the positioning of semi-fixed routes with extensions of parallel routes and zonal express to cover more scenarios.

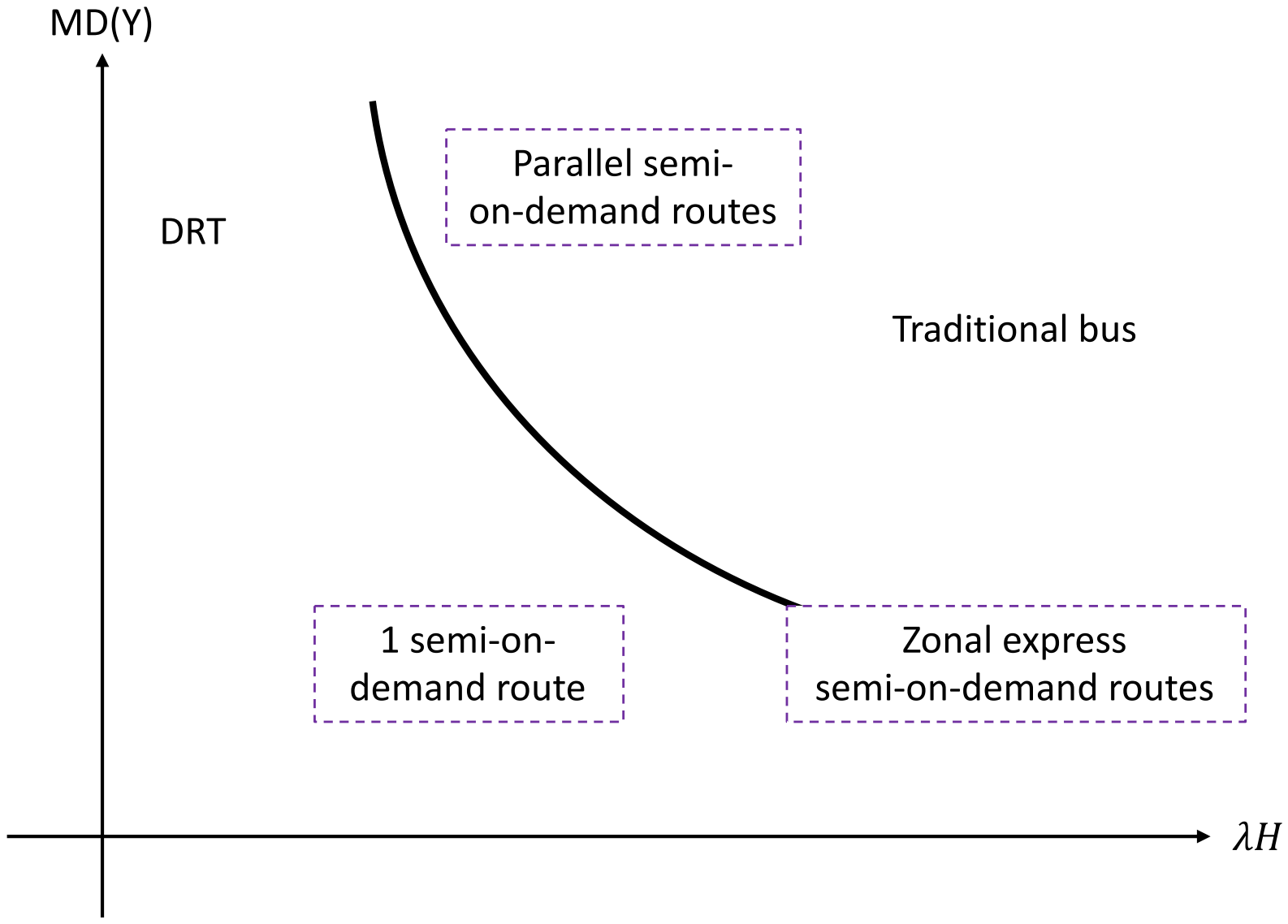

**Figure 7 - Illustration of Service Positioning**

## SIMULATION

Monte Carlo simulation was implemented with Python 3.8.8 to validate the benefits of the AMSoD service over fixed routes and the selection indicators derived in mathematical formulation. Results of general cases for single semi-on-demand and parallel routes are first presented, followed by case study with bus routes in Chicago.

Parameters values, summarized in **Table 3**, are taken with reference to previous studies (*5*, *13*, *27*). The value of time is evaluated as 50% of hourly median household income in 2019 in the U.S. (*28*, *29*). Existing bus stop spacing is set as 400m based on the practice of 2-5 stops per mile (*30*).

**Table 3 - Model Parameters**

| Category | Notation | Description | Model 1 | Model 2 | CTA #126 | CTA #84 |
|---|---|---|---|---|---|---|
| Cost | $\gamma_a$ | Penalty multiplier of access time | 2 | | | |
| | $\gamma_w$ | Penalty multiplier of waiting time | 1.5 | | | |
| | $\gamma_r$ | Penalty multiplier of riding time | 1 | | | |
| | $\gamma_o$ | Unit cost of vehicle distance | \$1/km | | | |
| | $VOT$ | Value of time | \$16.5/hour | | | |
| Distance | $L_x$ | x-direction length of a grid | 0.2km | | Various | |

<table>
<tr><th>Category</th><th>Notation</th><th>Description</th><th>Model 1</th><th>Model 2</th><th>CTA #126</th><th>CTA #84</th></tr>
<tr><td rowspan="3"></td><td>$L_y$</td><td>y-direction length of a grid</td><td colspan="2">0.1km</td><td colspan="2">Various</td></tr>
<tr><td>$GL_x$</td><td>x-direction suburb length</td><td colspan="2">10km</td><td>10.9km</td><td>8km</td></tr>
<tr><td>$GL_y$</td><td>y-direction max. coverage width</td><td>0.53km</td><td>2km</td><td colspan="2">Various</td></tr>
<tr><td rowspan="2">Time</td><td>$t_s$</td><td>Stopping time per stop of fixed-route bus service</td><td colspan="2">0.4min</td><td colspan="2">0.33min</td></tr>
<tr><td>$t_s'$</td><td>Stopping time per pick-up of AMSoD service</td><td colspan="4">0.4min</td></tr>
<tr><td rowspan="2">Speed</td><td>$v_w$</td><td>Average walking speed of travelers</td><td colspan="4">4km/h</td></tr>
<tr><td>$v_d$</td><td>Average speed of buses in suburb section (excluding stopping for boarding / alighting)</td><td colspan="2">35km/h</td><td colspan="2">30km/h</td></tr>
<tr><td rowspan="2">Passenger</td><td>$\lambda_K$</td><td>Number of passengers per hour</td><td colspan="2">60</td><td>80</td><td>50</td></tr>
<tr><td>$s_o$</td><td>Maximum access time for passengers to take the bus</td><td>8min</td><td>30min</td><td colspan="2">Various</td></tr>
<tr><td rowspan="4">Bus</td><td>$d_{xs}$</td><td>Fixed-route station spacing in x-direction</td><td colspan="2">400m</td><td colspan="2">Various</td></tr>
<tr><td>$H$</td><td>Headway (departure for AMSoD)</td><td colspan="3">15min</td><td>20min</td></tr>
<tr><td>$Cap$</td><td>Bus capacity</td><td colspan="4">30</td></tr>
<tr><td>$n_p$</td><td>Number of parallel routes</td><td>1</td><td>2</td><td colspan="2">1</td></tr>
<tr><td rowspan="3">Model</td><td>$J$</td><td>Number of trips run in the main model</td><td colspan="4">10000</td></tr>
<tr><td>$J_s$</td><td>Number of trips run in each model of sensitivity analysis</td><td colspan="4">1000</td></tr>
<tr><td>$T$</td><td>Time duration in the simulation</td><td colspan="4">3hours</td></tr>
</table>

Demand is allocated to the coverage area of traditional bus stops. The spatial distribution of demand points is uniform from the existing bus stop in x- and y-directions (Eq. (39) and (40)), within the maximum access distance defined (Eq. (41)).

$$x_{jk} \sim U\left(x_{bs,bs_k} - \frac{d_x}{2}, x_{bs,bs_k} + \frac{d_x}{2}\right) \quad (39)$$

$$y_{jk} \sim U(-GL_y, GL_y) \tag{40}$$

$$|x_{bs,bs_k} - x_{j,k}| + |y_{j,k}| \leq GL_y \tag{41}$$

Minibuses run on rules by previous assumptions. From a request point, a bus travels to the next intersection and turns by y-direction and then x-direction to the next request point. For two request points in the same x-grid, it travels in a loop if necessary to pick up both passengers.

In the implementation, all requests are sorted by x-coordinates, then checked one by one to confirm they are ready for pick up in front of the bus. The minibus then moves forward to the request point and looks for the next passenger through the same process. A bus capacity is set such that unserved passengers are picked up on the next trip. Passengers departing in the second hour are counted for cost calculation, as initial demands may experience unreasonable waiting time and demand may be unsatisfied by the last trips.

Detailed cost components with median and 95% confidence intervals (in brackets) are tabulated in **Table 4**. **Figure 8** shows examples of bus trips where $K$ represents the number of passengers served, $T$ the journey period (in hours), $t$ the suburb IVTT (in hours), green dots the traditional bus stops, and blue dots the request points.

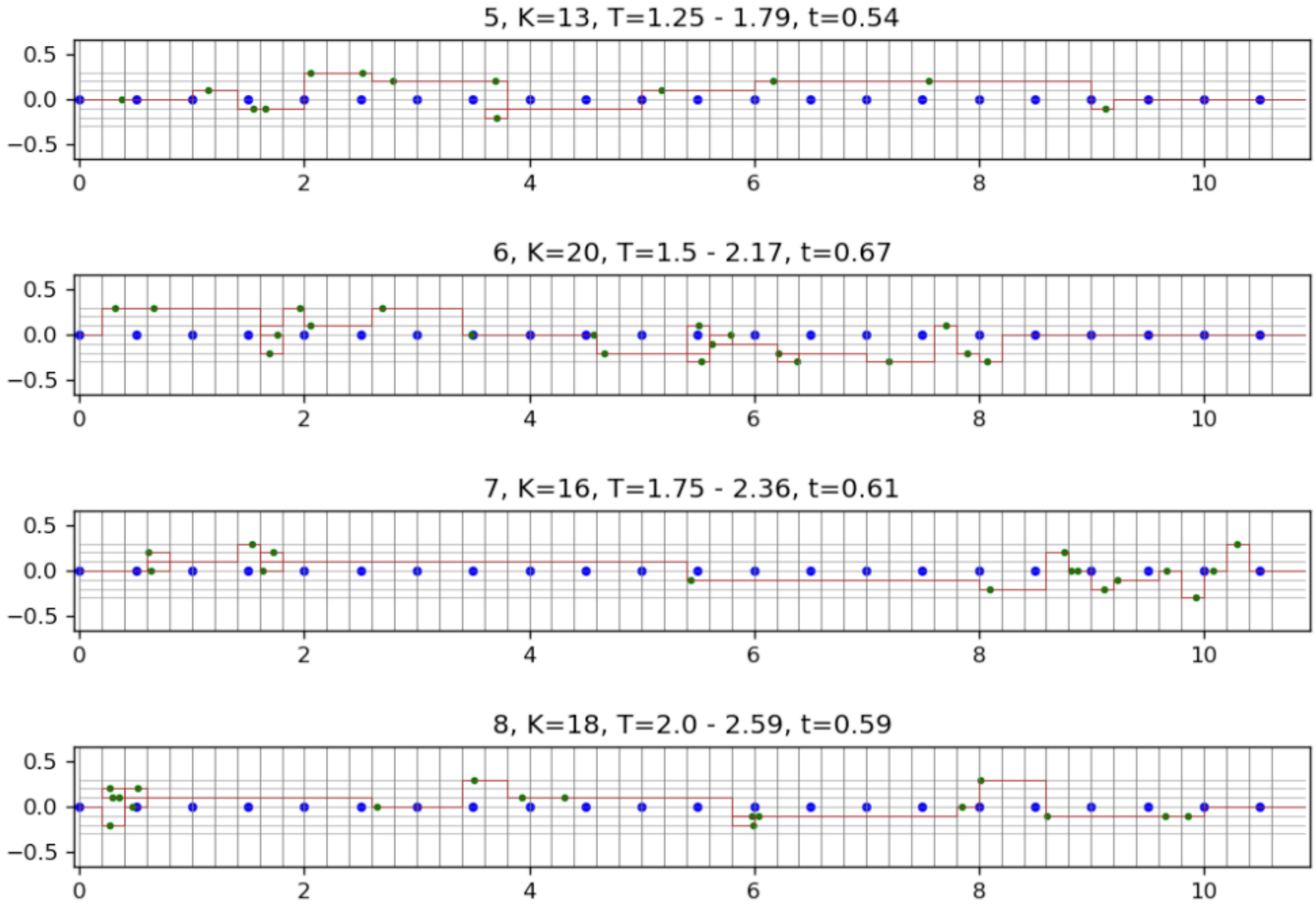

**Figure 8 - Examples of Bus Trips**

**Table 4 - Medians and Confidence Intervals of Model Results**

| **Parameter** | **Model 1** | | **Model 2** | | **CTA Bus #126** | | **CTA Bus #84** | |
|---|---|---|---|---|---|---|---|---|
| | **Fixed Route** | **AMSoD Route** | **Fixed Route** | **2 Parallel AMSoD Routes** | **Fixed Route** | **AMSoD Route** | **Fixed Route** | **AMSoD Route** |
| Average waiting time, $\overline{s_{wk}}$ (min) | 7.5 (6.4 – 8.6) | 8.4 (6.6 – 11.2) | 7.5 (6.4 – 8.6) | 15.9 (12.6 – 20.6) | 7.5 (6.6 – 8.5) | 8.6 (6.9 – 11.4) | 10.0 (8.4 – 11.6) | 11.2 (8.6 – 15.4) |
| Average IVTT (min) | 13.9 (12.2 – 15.6) | 16.1 (13.5 – 19.4) | 13.9 (12.1 – 15.6) | 21.7 (17.8 – 26.8) | 22.1 (20.0 – 24.2) | 19.2 (16.8 – 21.8) | 10.5 (8.9 – 12.0) | 17.0 (13.3 – 21.5) |
| Access cost, $C_a$ ($) | 158 (121 – 196) | 0 | 592 (458 – 738) | 0 | 214 (171 – 260) | 0 | 233 (176 – 296) | 0 |
| Waiting cost, $C_w$ ($) | 186 (141 – 235) | 208 (146 – 302) | 185 (140 – 232) | 392 (275 – 561) | 248 (196 – 304) | 283 (205 – 409) | 211 (155 – 271) | 238 (162 – 359) |
| Riding cost, $C_r$ ($) | 229 (175 – 285) | 267 (187 – 361) | 229 (176 – 283) | 358 (253 – 494) | 488 (391 – 586) | 425 (321 – 540) | 147 (109 – 187) | 239 (157 – 342) |
| Operator's cost, $C_o$ ($) | 120 | 163 (127 – 218) | 120 | 216 (149 – 310) | 131 | 170 (139 – 217) | 72 | 85 (72 – 112) |
| Generalized cost, $\sum TC$ ($) | 693 (570 – 820) | 642 (503 – 824) | 1126 (914 – 1350) | 970 (748 – 1264) | 1082 (905 – 1260) | 882 (705 – 1104) | 664 (527 – 808) | 563 (409 – 776) |
| Generalized cost difference, $\Delta TC$ ($) | -51 (-134 – 67) | | -154 (-312 – 43) | | -197 (-299 – -67) | | -99 (-193 – 38) | |
| Number of passengers included | 60 (48 – 72) | | 60 (48 – 73) | | 80 (66 – 95) | | 51 (40 – 63) | |

**Model 1 - Single Semi-on-Demand Route**

Grid spacing is taken as 200m x 100m with a 10km length to reflect a typical suburb setting. The y-directional width coverage area, $GL_y = v_w s_o = 4 \times \frac{8}{60} = 0.53\ km$. By Eq. (34) and $MD(Y) = \frac{2}{3} GL_y = 0.36, SI = 0.80 < 1$, indicating that the AMSoD service is beneficial.

*Modeling Results*

The average waiting time of AMSoD route is around 1 min longer than the fixed route and the average IVTT increases by around 2 min. This aligns with previous discussion that randomness due to y-directional detour and number of stops increases expected waiting time. By the tradeoff between saving in access cost, and extra waiting, riding, and operator's costs, AMSoD route gives a lower generalized cost, which amounts to an average of $0.9 per passenger and is roughly skewed to the left compared to the normal distribution line in **Figure 9**.

A significant portion of AMSoD service originated from the saving in access cost, which were evaluated based on assumed values of access time. If we conservatively consider the lower bound of plausible range recommended in (*29*), the saving would be reduced by 20% but the difference in generalized cost would still be negative, favoring the new service. It should also be noted that under such assumption, other values of time should be lower as well, further supporting the finding.

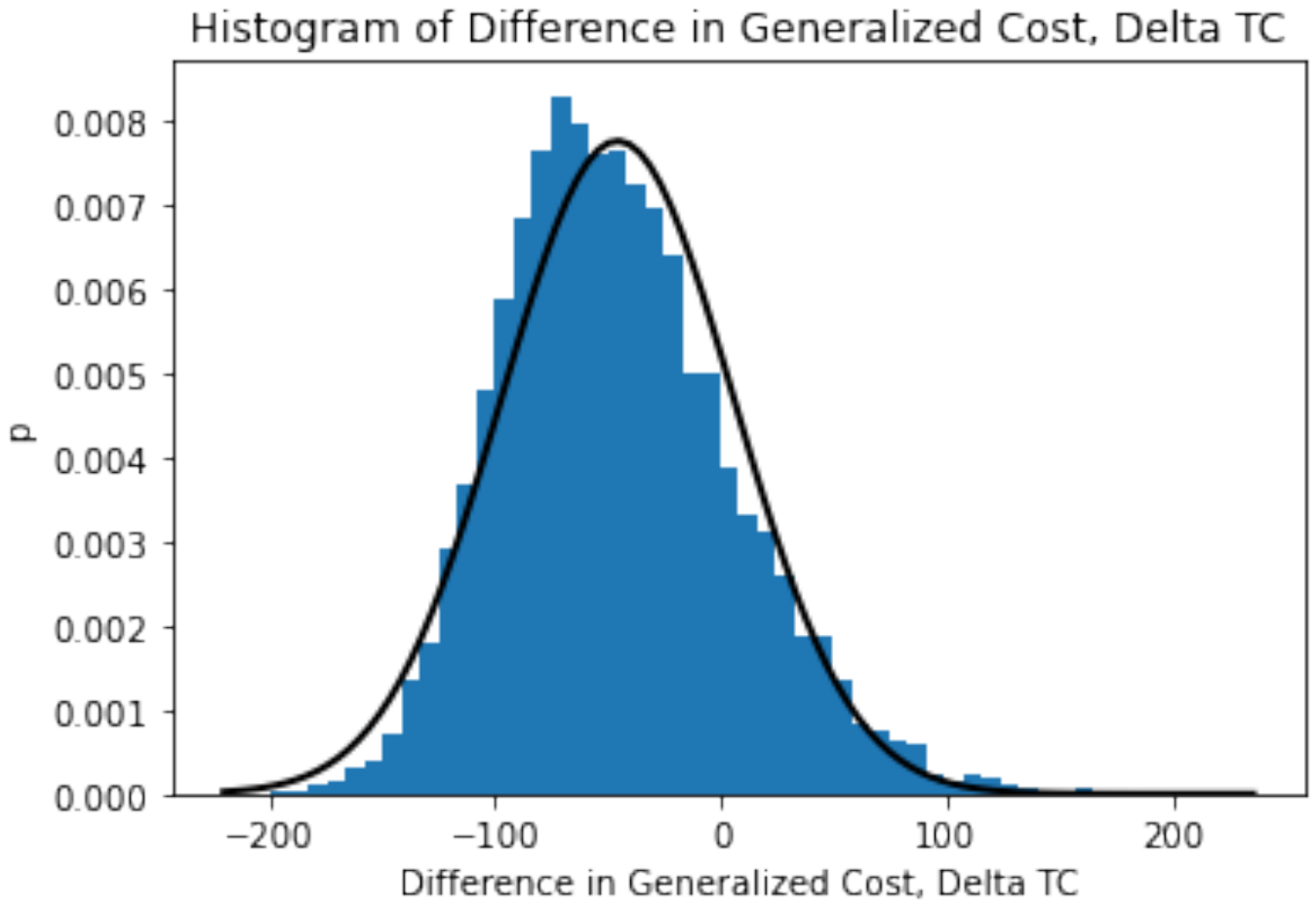

**Figure 9 - Histogram of Sum of Differences in Generalized Cost, $\sum \Delta TC$**

*Sensitivity Analysis on Bus Capacity*

In **Figure 10,** there is a sharp decrease in difference in generalized cost when the bus capacity increases from 15 to 20, contributed by lower waiting time further from near-capacity. The IVTT also increases slightly for more detours when a bus can carry more passengers in rare cases (**Figure 11**). At capacities higher than 20, the generalized costs level off. The slight decrease in waiting time may not justify deploying buses of larger sizes.

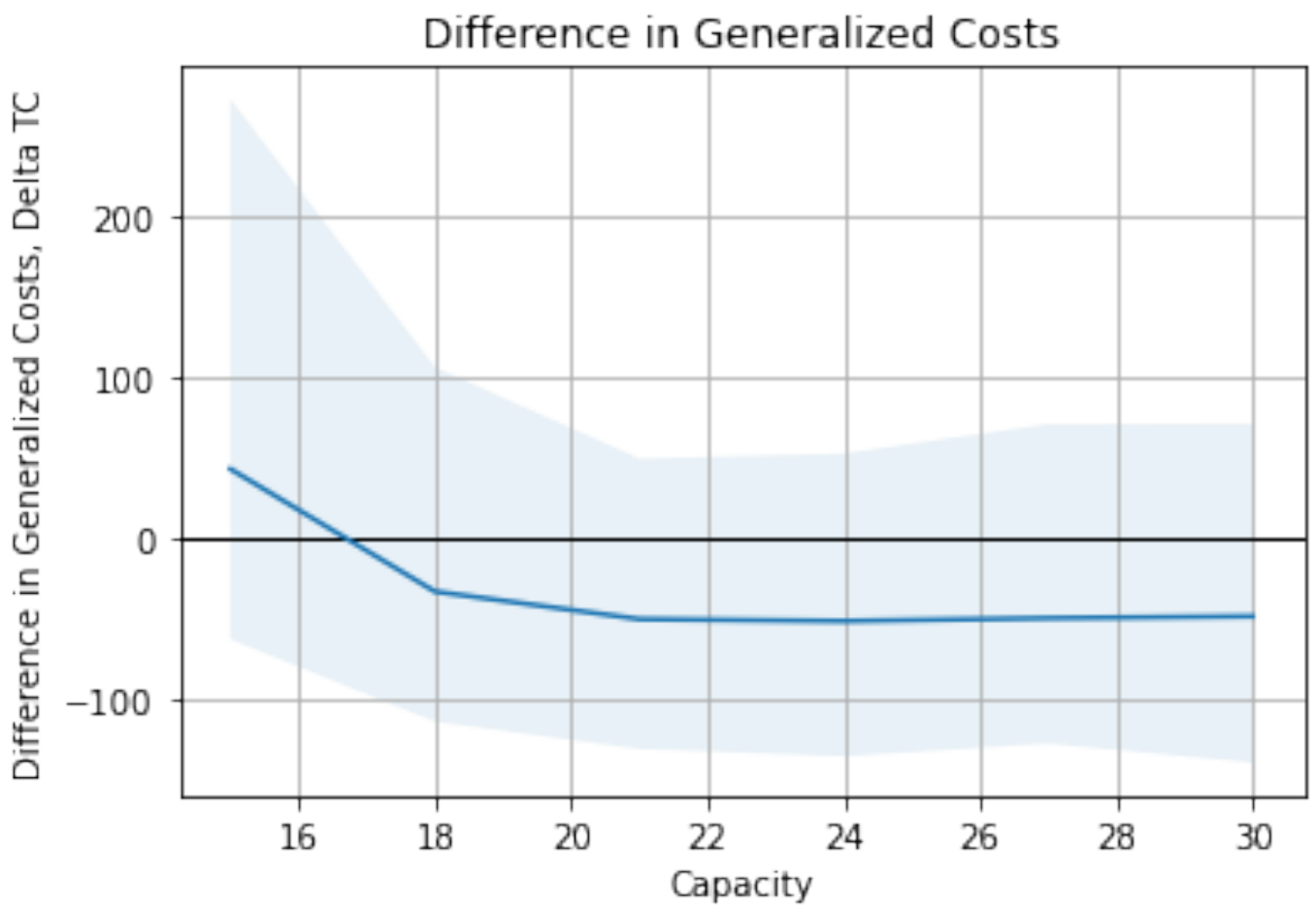

**Figure 10 - Difference in Generalized Costs at Different Capacities**

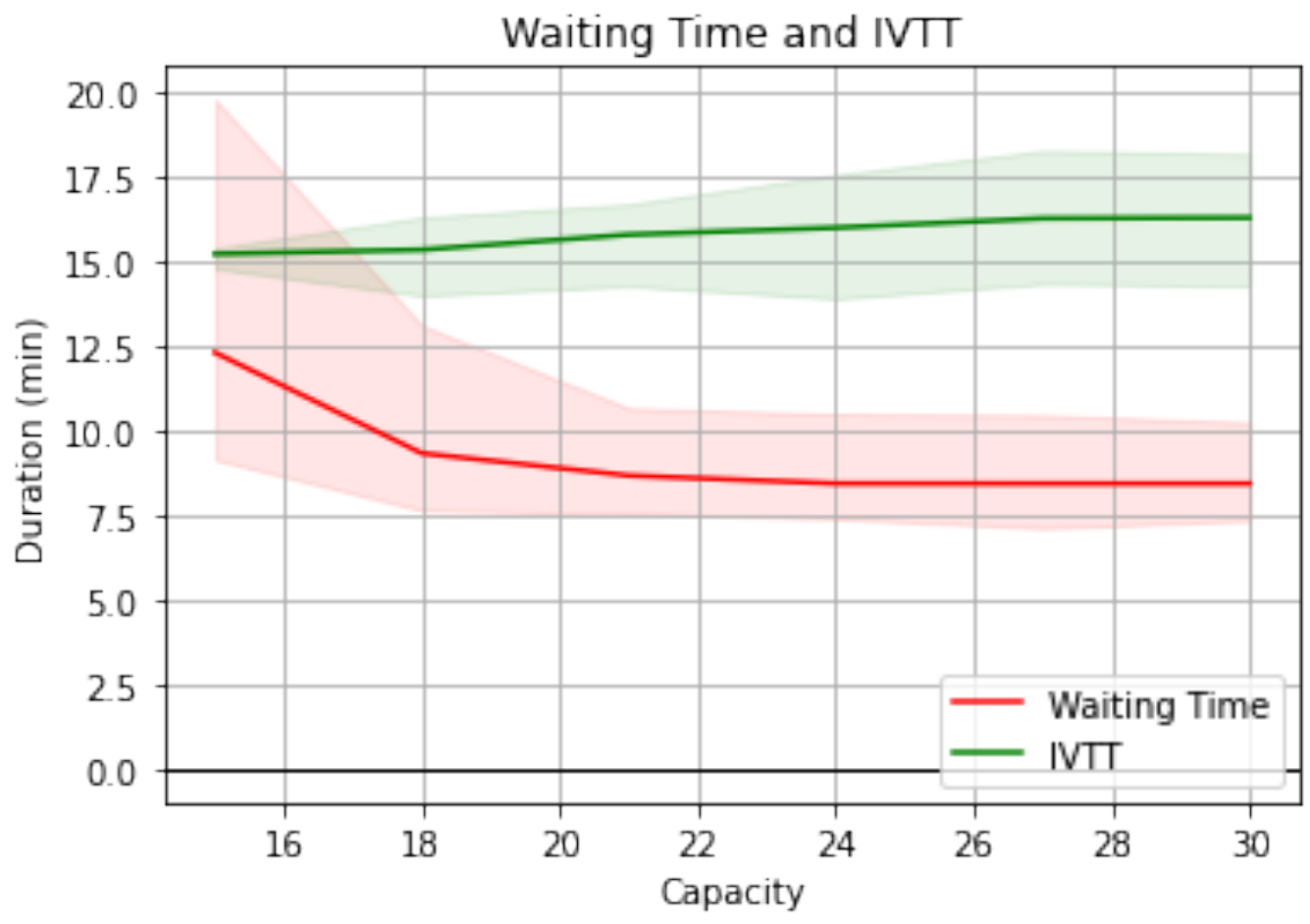

**Figure 11 - Mean Waiting Time and IVTT at Different Capacities**

*Sensitivity Analysis on Demand*

Different hourly demand flows are tested to explore the impact of demand on the AMSoD benefits. In **Figure 12**, the difference in generalized costs increases from 40 passengers/hour and reaches positive before 80 passengers/hour. From **Figure 13**, this is due to the increases in IVTT and waiting time for extra y-directional detours. The result is close to the results of Eq. (35), $\lambda_k \leq 88$, where the discrepancy can be explained by the ignored saving in IVTT from the reduced number of stops.

This suggests lower demand favor AMSoD routes and echoes the development of pioneering autonomous shuttles and smaller size in DRT.

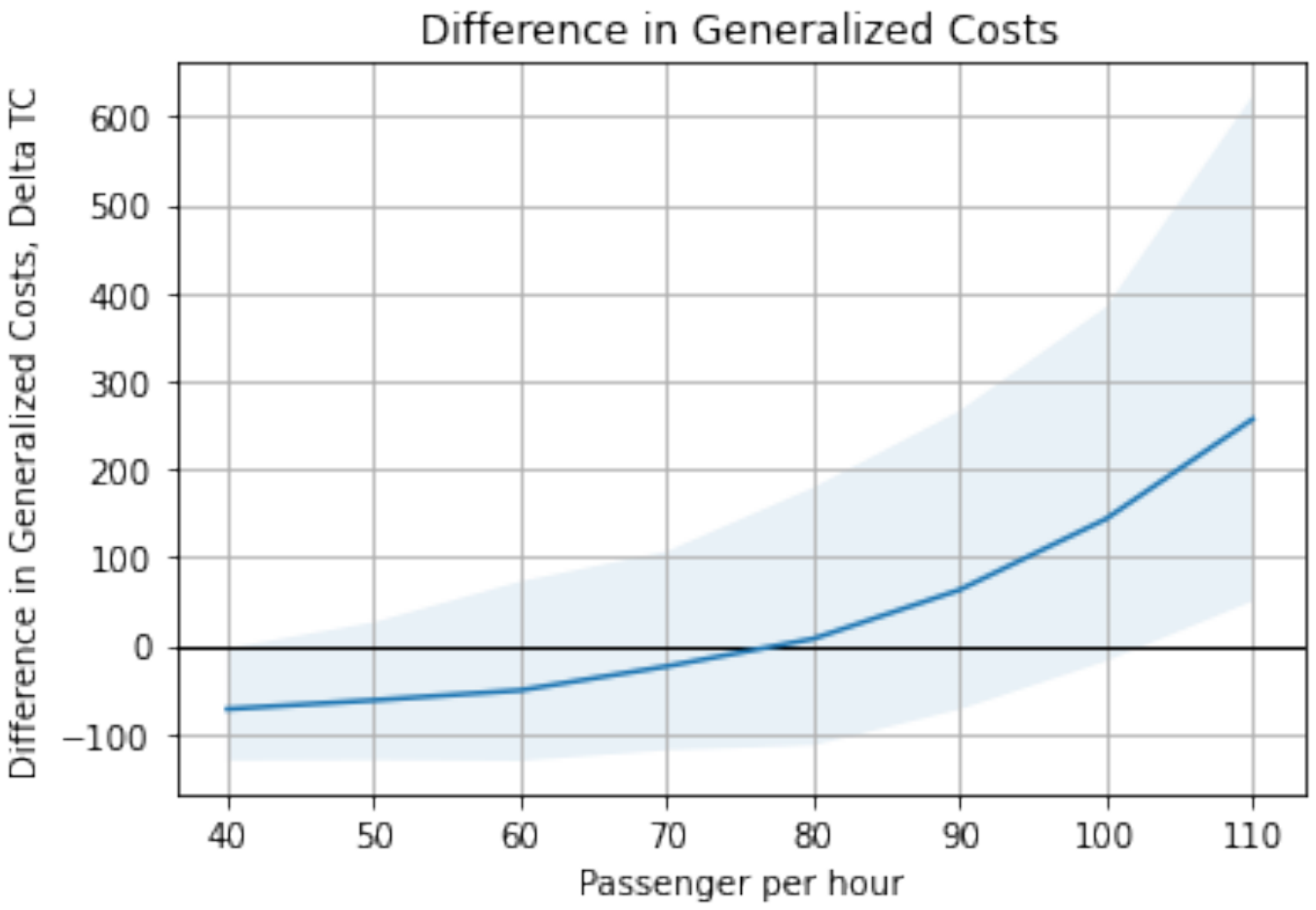

**Figure 12 - Difference in Generalized Costs at Different Hourly Demand**

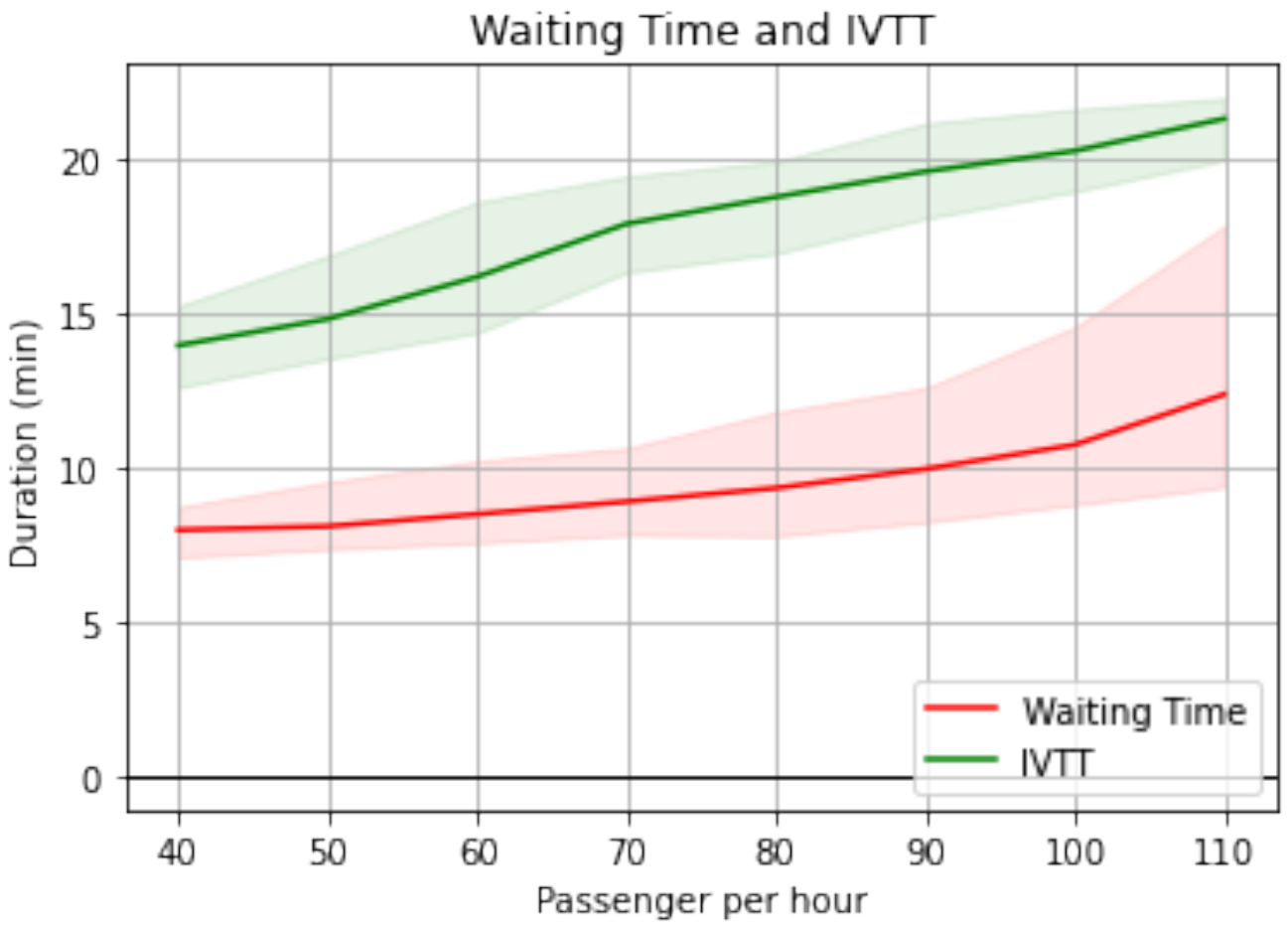

**Figure 13 - Mean Waiting Time and IVTT at Different Hourly Demand**

**Model 2 - Parallel Semi-on-Demand Routes**

To showcase the benefits for parallel routes, Model 2 utilized a similar case to model 1 except that the coverage area is larger. The maximum access time is 30min with $MD(Y) = 1.33$, posing inefficiencies to semi-on-demand routes shown by its $SI = 0.97$. This can be better served by two parallel semi-on-demand routes, each covering half of the service area, as shown by $SI_p = 0.88$ in Eq. (36).

*Model Results*

In Figure 14, the first bus trip picks up the passengers in the lower portion of the service area, while the second bus trip covers the upper portion. This brings a significant reduction in y-directional travel.

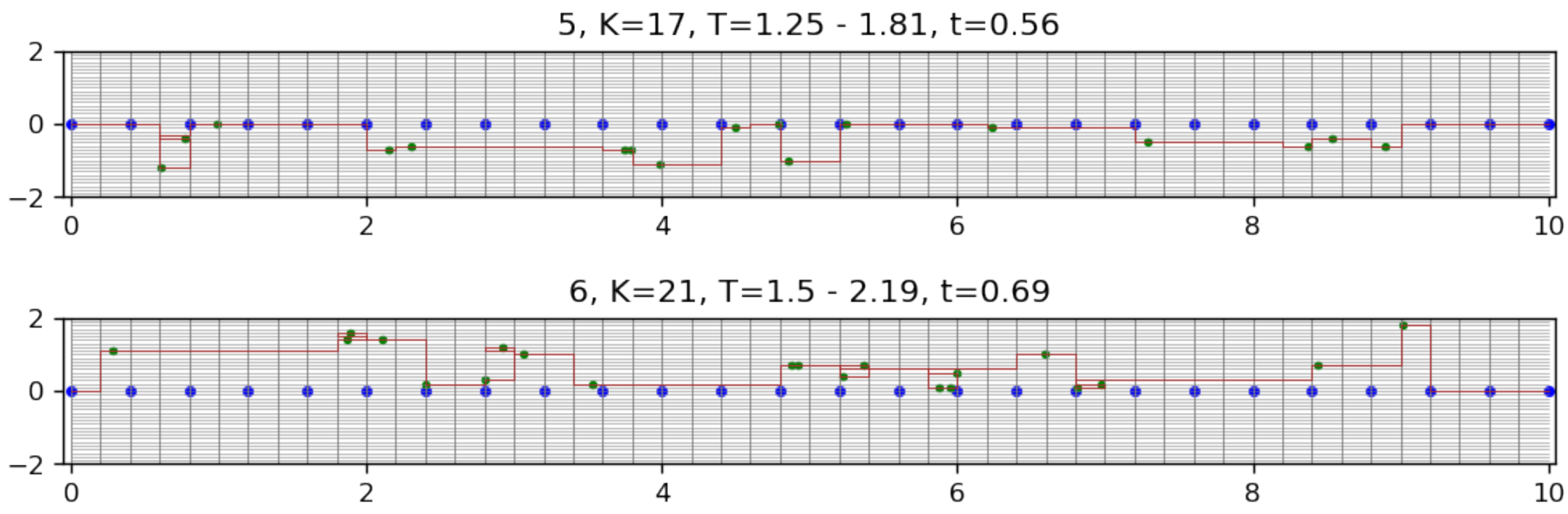

**Figure 14 - Examples of Bus Trips in Model 2 Results**

The average waiting time of semi-on-demand routes is more than double that of fixed routes due to split frequency, while IVTT also increases by nearly 8 min due to considerable detours for pick-ups. This however cut access cost and generalized costs (**Figure 15**).

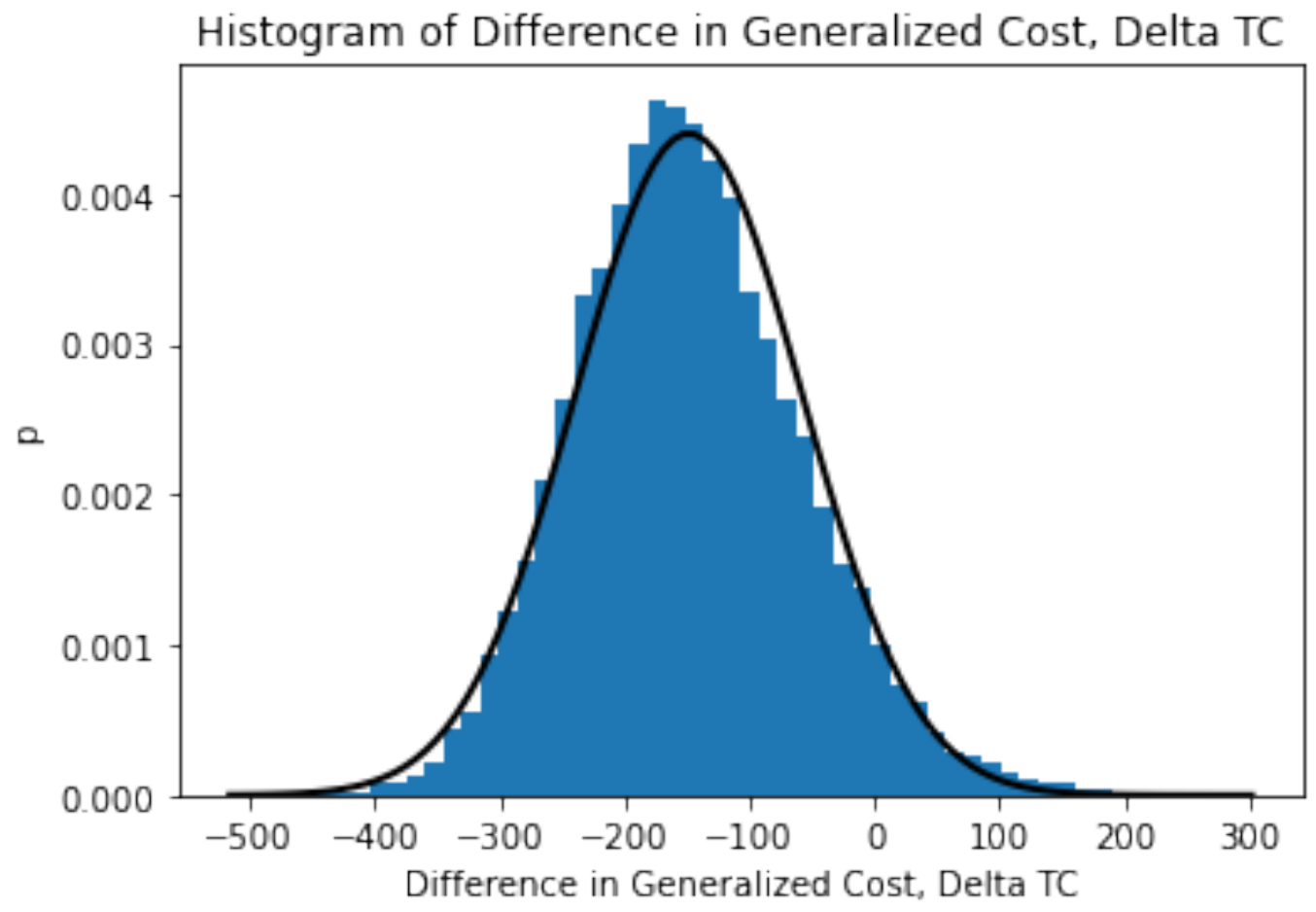

**Figure 15 - Histogram of Sum of Differences in Generalized Cost, $\sum \Delta TC$**

*Sensitivity Analysis on Bus Capacity*

The results are similar to Model 1, where generalized costs level off after adequate capacities (*i.e.* 20/bus) (**Figure 16 - Figure 18**).

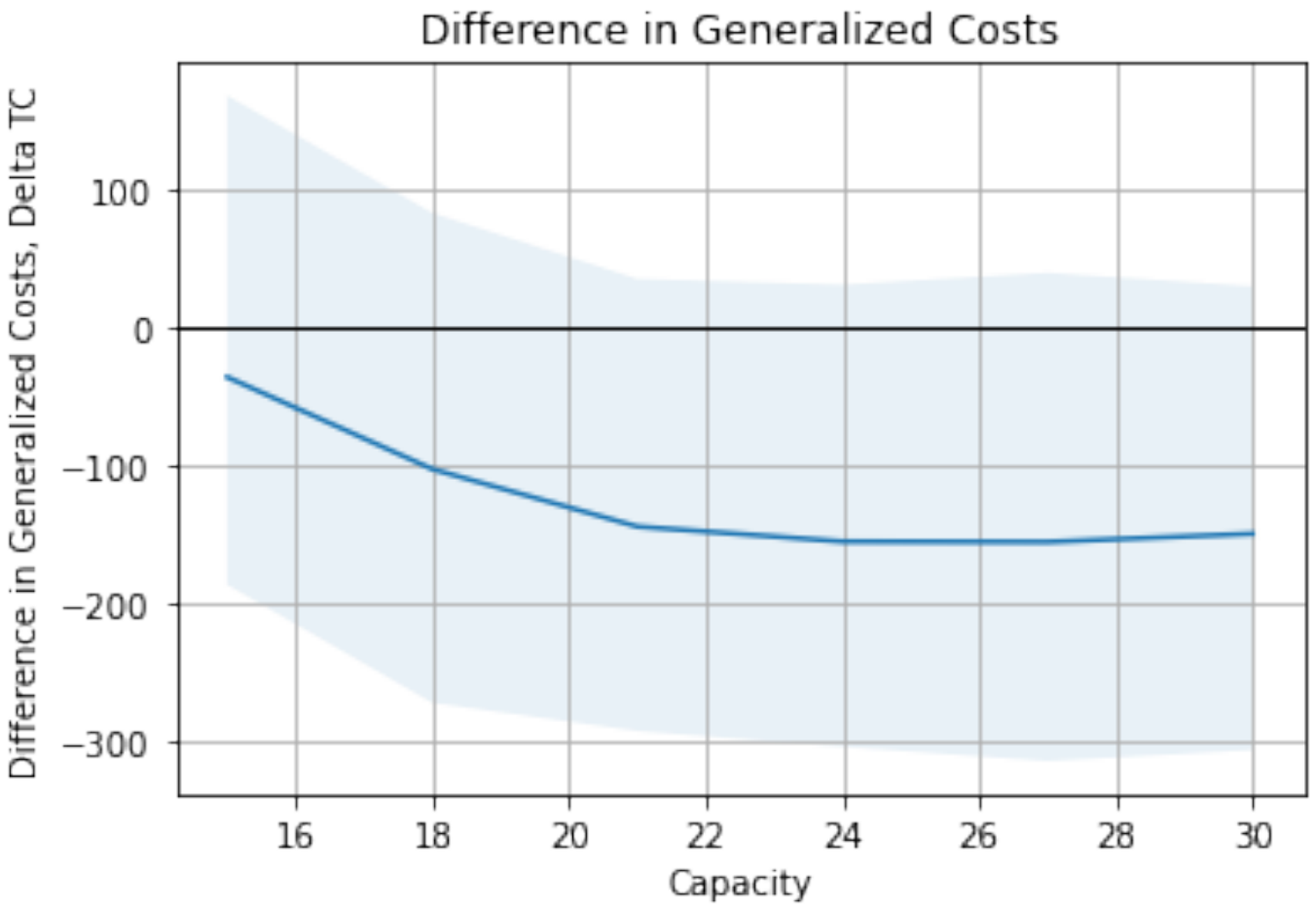

**Figure 16 - Difference in Generalized Costs at Different Capacities**

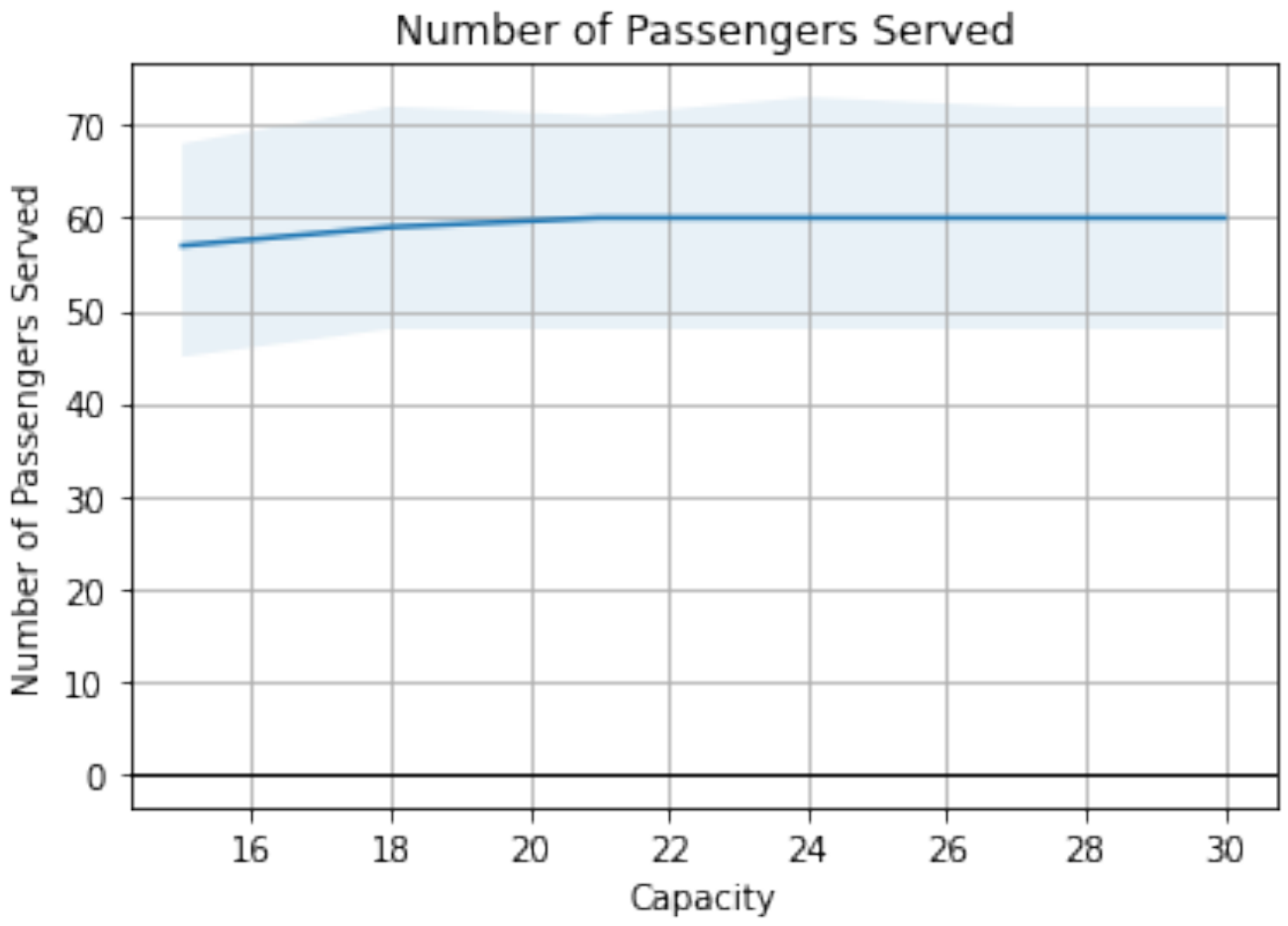

**Figure 17 - Number of Passengers Served at Different Capacities**

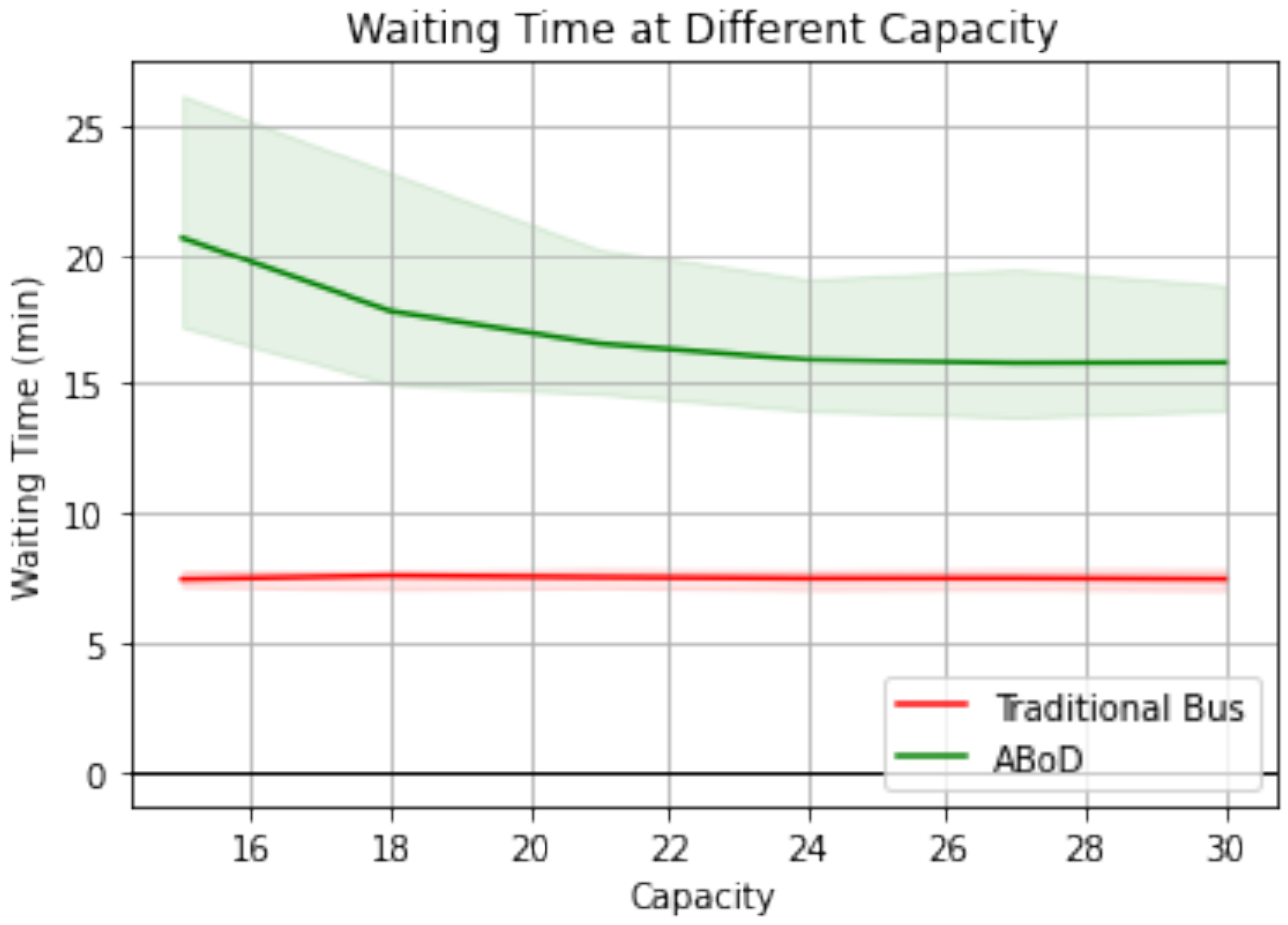

**Figure 18 - Mean Waiting Time and IVTT at Different Capacities**

*Sensitivity Analysis on Demand*

**Figure 19** shows generalized costs difference is the lowest at around 60 passengers/hour and remain negative until 90 passengers/hour, which agrees with Eq. (38), $\lambda_k \leq 97$. This is also brought by the increase in waiting time and IVTT at high demand (**Figure 20**).

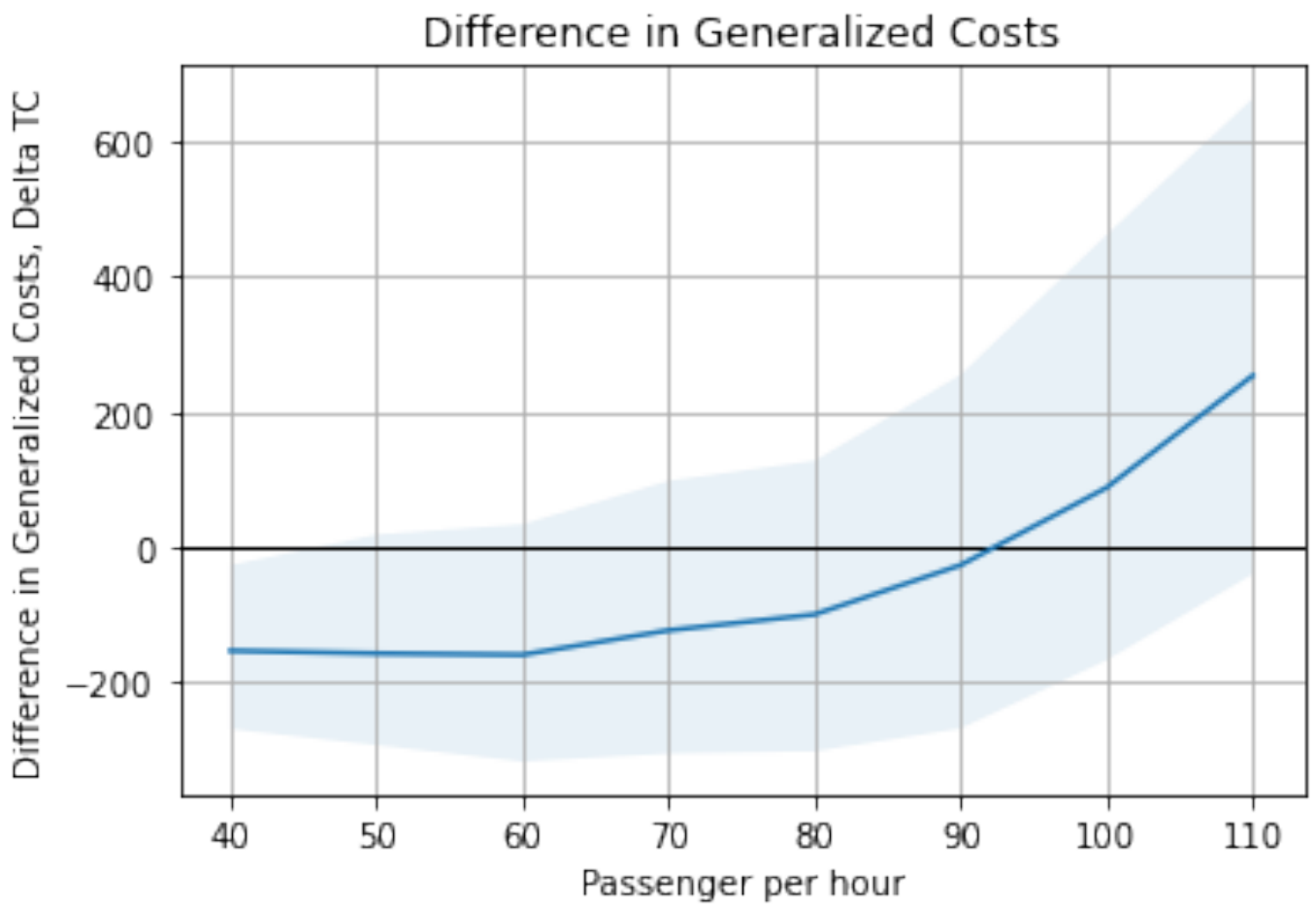

**Figure 19 - Difference in Generalized Costs at Different Hourly Demand**

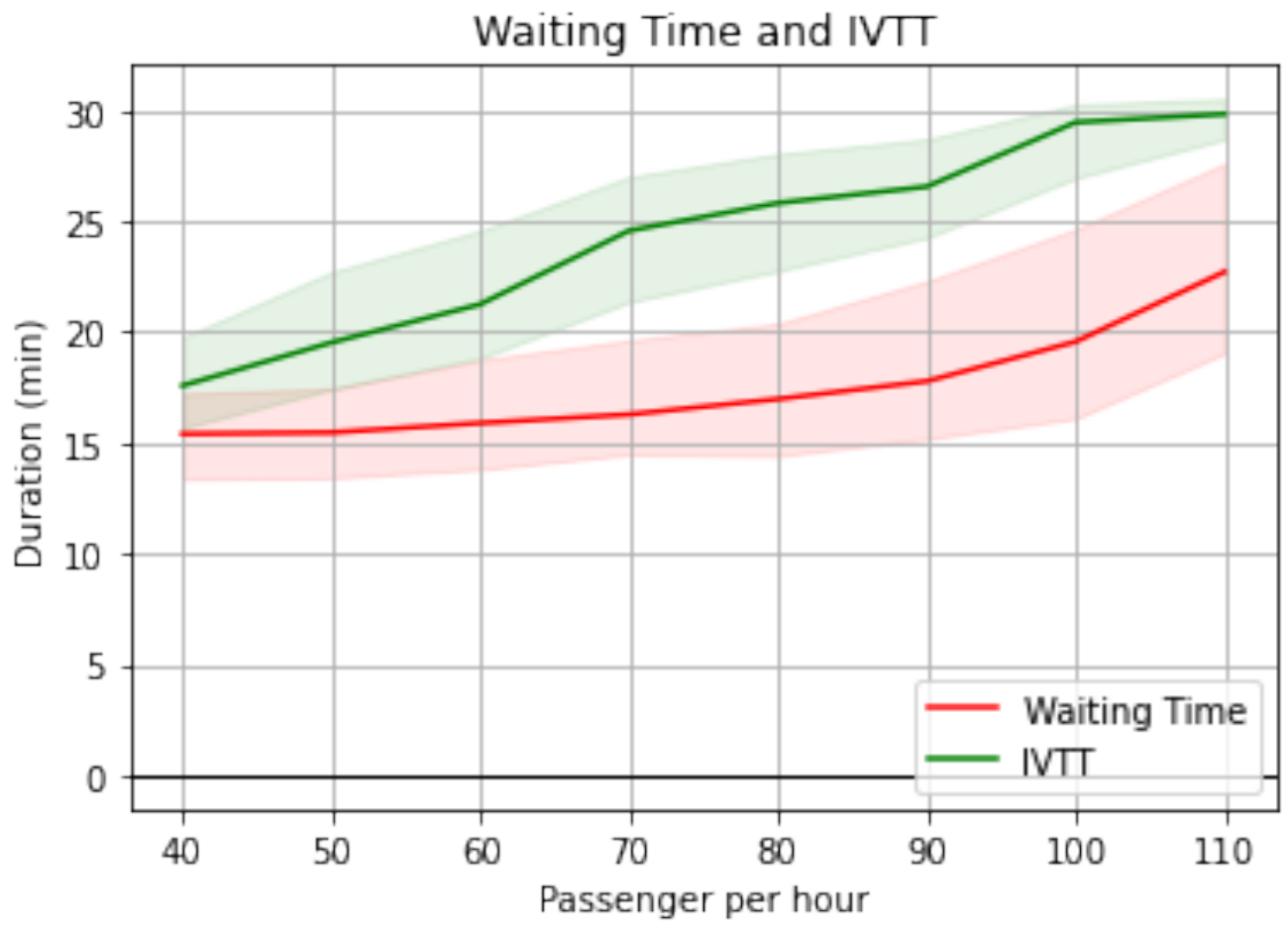

**Figure 20 - Mean Waiting Time and IVTT at Different Hourly Demand**

**CTA Bus #126**

To demonstrate the advantages of AMSoD service in the real world, case studies on bus routes in Chicago are carried out with open data from Chicago Transit Authority (CTA) including bus boarding by stop and schedule (*31*). While the exact demand points are not available, uniform distribution within an access distance is assumed pro-rata along existing bus stops based on the average weekday bus stop boardings in October 2012.

The main service of CTA Bus #126, which travels eastbound along West Jackson Boulevard towards the Loop, was chosen as an example. The parallel routes #20 and #7 and the Blue Line of "L" metro limit its catchment area to be 200m on both sides (**Figure 21**). Periods of 9 a.m. – 2:30 p.m. and 5:30 p.m. – 10:30 p.m. are served with headway not smaller than 15 min, indicating a perceived passenger flow of around 80/hour (*32*).

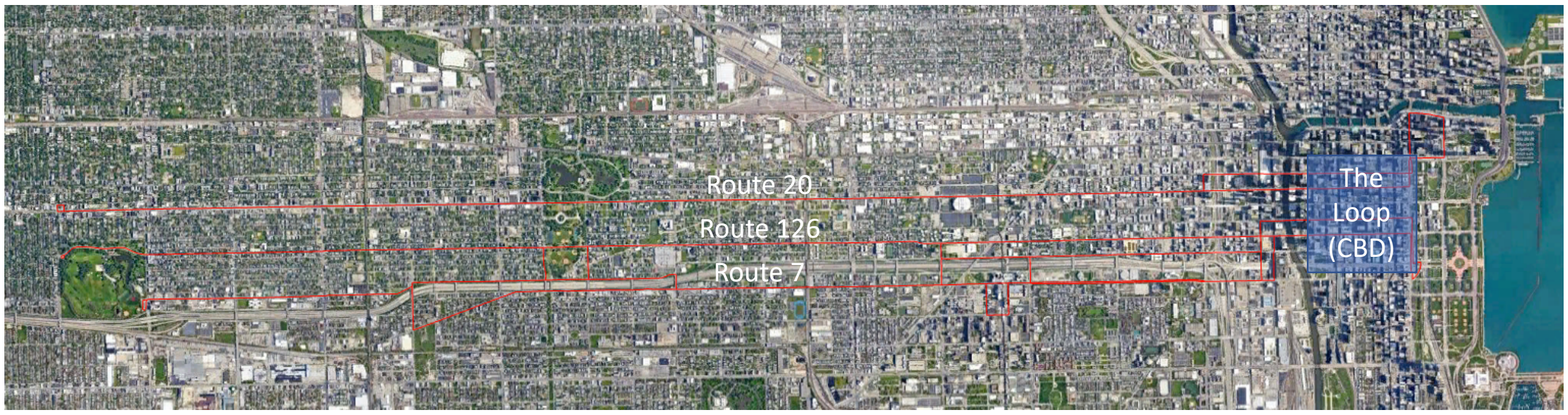

**Figure 21 - Map of CTA Bus #126 (*33–35*)**

As shown in **Figure 22**, the road network in the catchment area is represented by grids, with blue dots for existing bus stops and eastward chainage in kilometers. The north-south displacement of the fixed route is negligible. By Eq. (34) and $MD(Y) = 0.2\left(\frac{2}{3}\right) = 0.133$, $SI = 0.75 < 1$, implying improvement brought by AMSoD route.

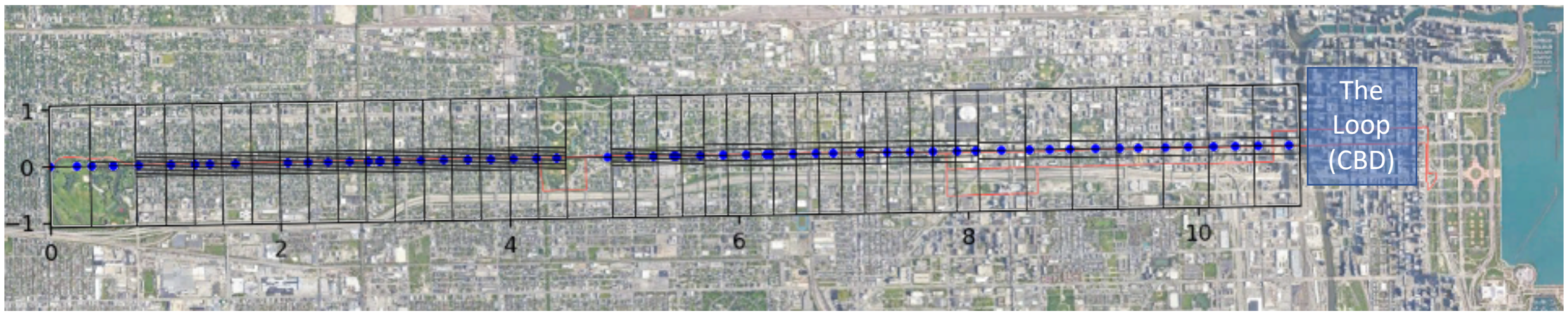

**Figure 22 - Grid model for CTA Bus #126**

*Modeling Result*

The average waiting time of AMSoD is around 1 min higher than that of the fixed route. In contrast, its average IVTT of AMSoD is around 3 min lower, because the bus stop density (5/km) is higher than the previous assumption (2.5/km), resulting in a larger scheduled dwell time saving. The difference in generalized cost is also skewed to the left (**Figure 23**), with a median of -197, equivalent to $2.5 benefit per passenger.

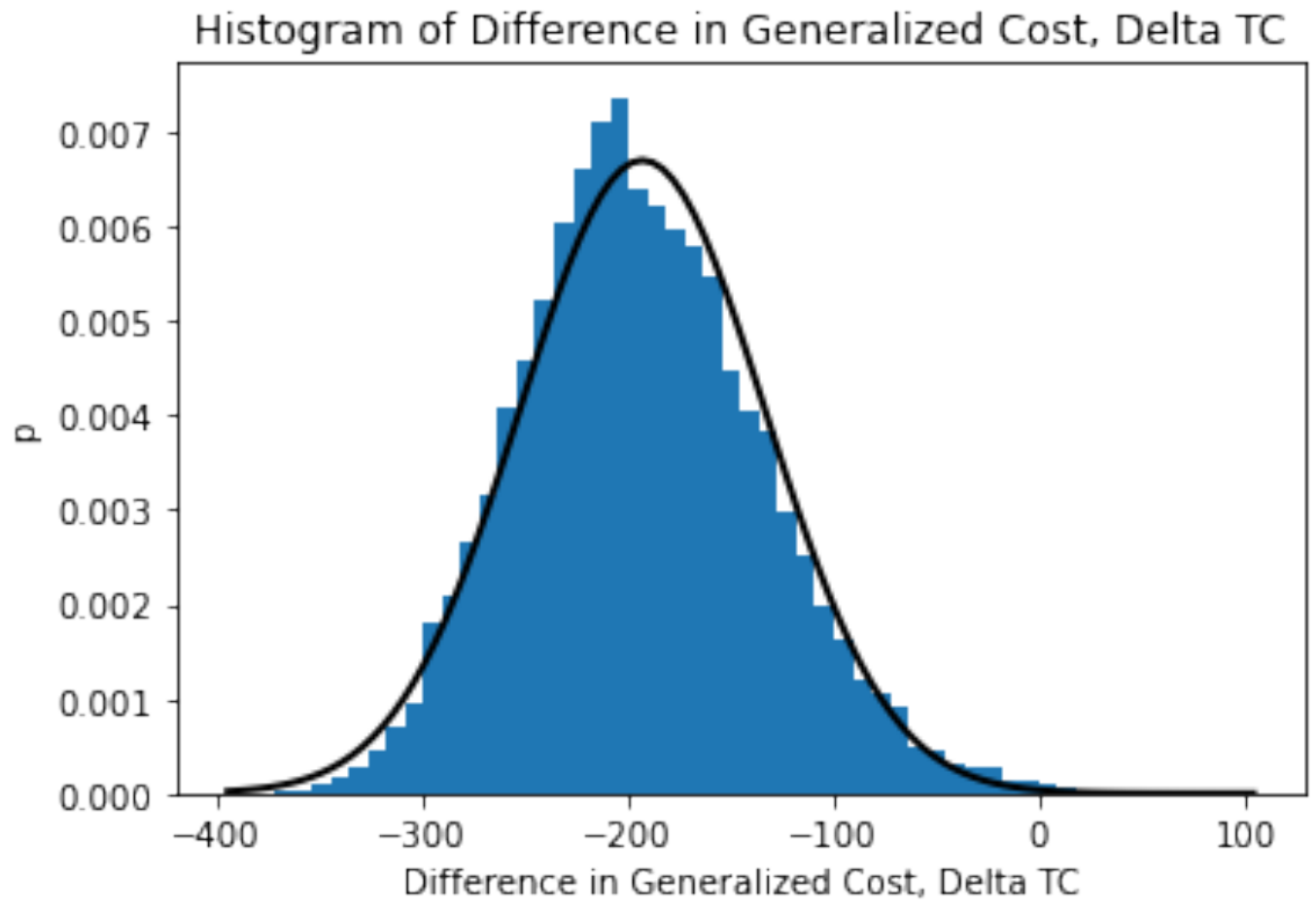

**Figure 23 - Histogram of Sum of Difference in Generalized Cost, $\sum \Delta TC$**

*Sensitivity Analysis on Bus Capacity*

**Figure 24** shows a similar trend to the previous model. Starting from an adequate capacity of 20 as shown in **Figure 25**, the benefits of increasing bus capacity to AMSoD service are manifested with reduced waiting time despite a slight increase in IVTT in **Figure 26**.

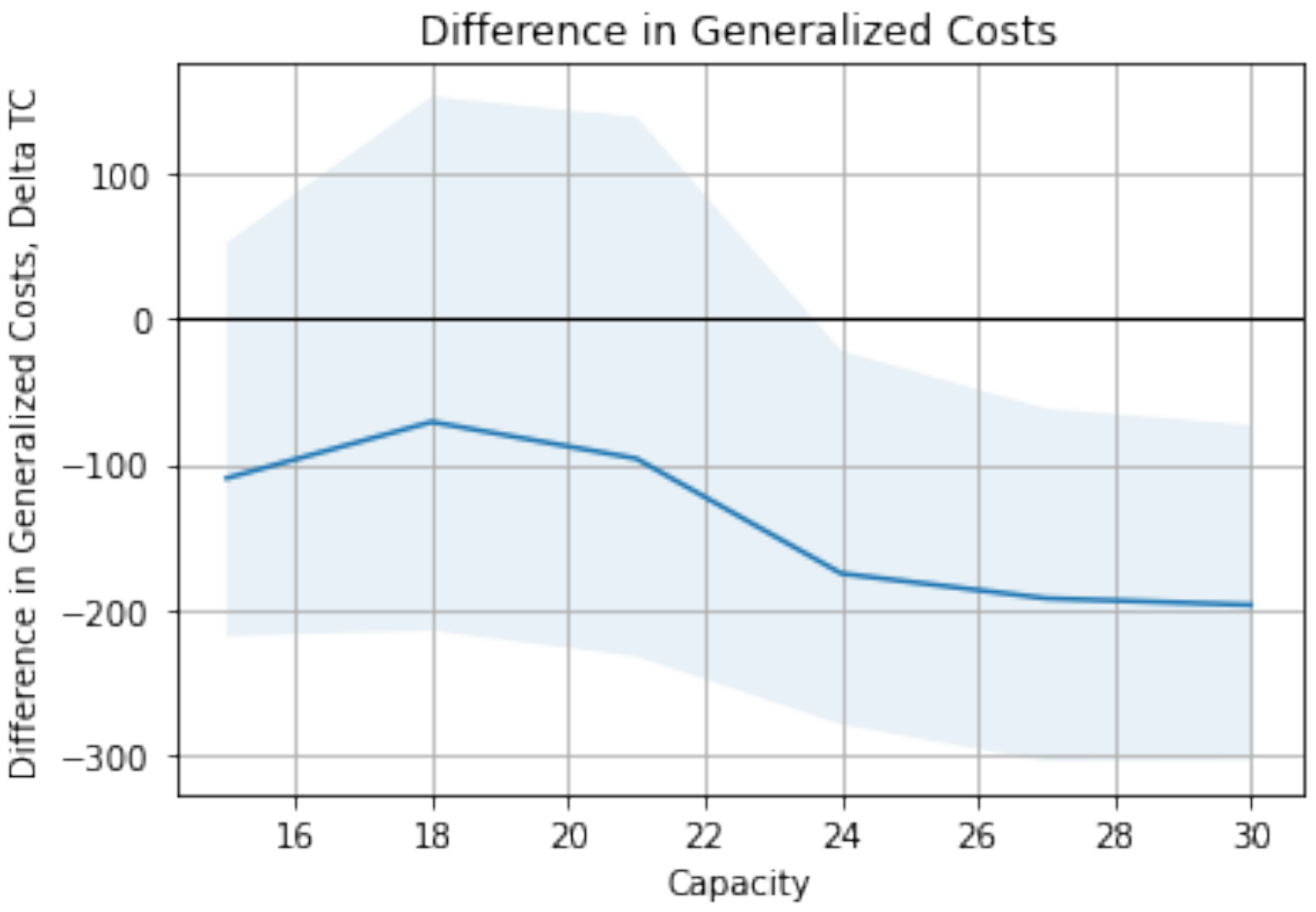

**Figure 24 - Difference in Generalized Costs at Different Capacities**

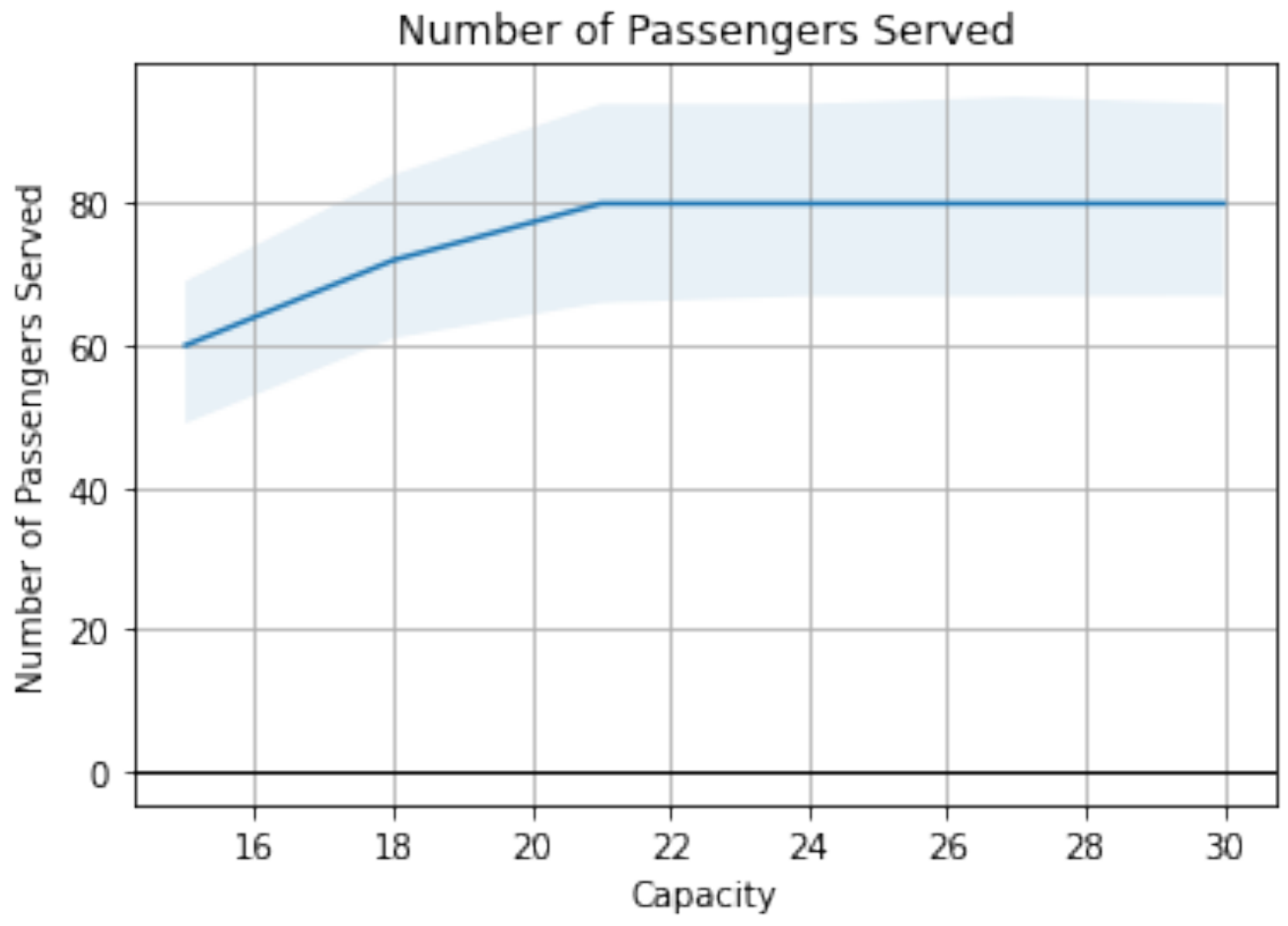

**Figure 25 - Number of Passengers Served at Different Capacities**

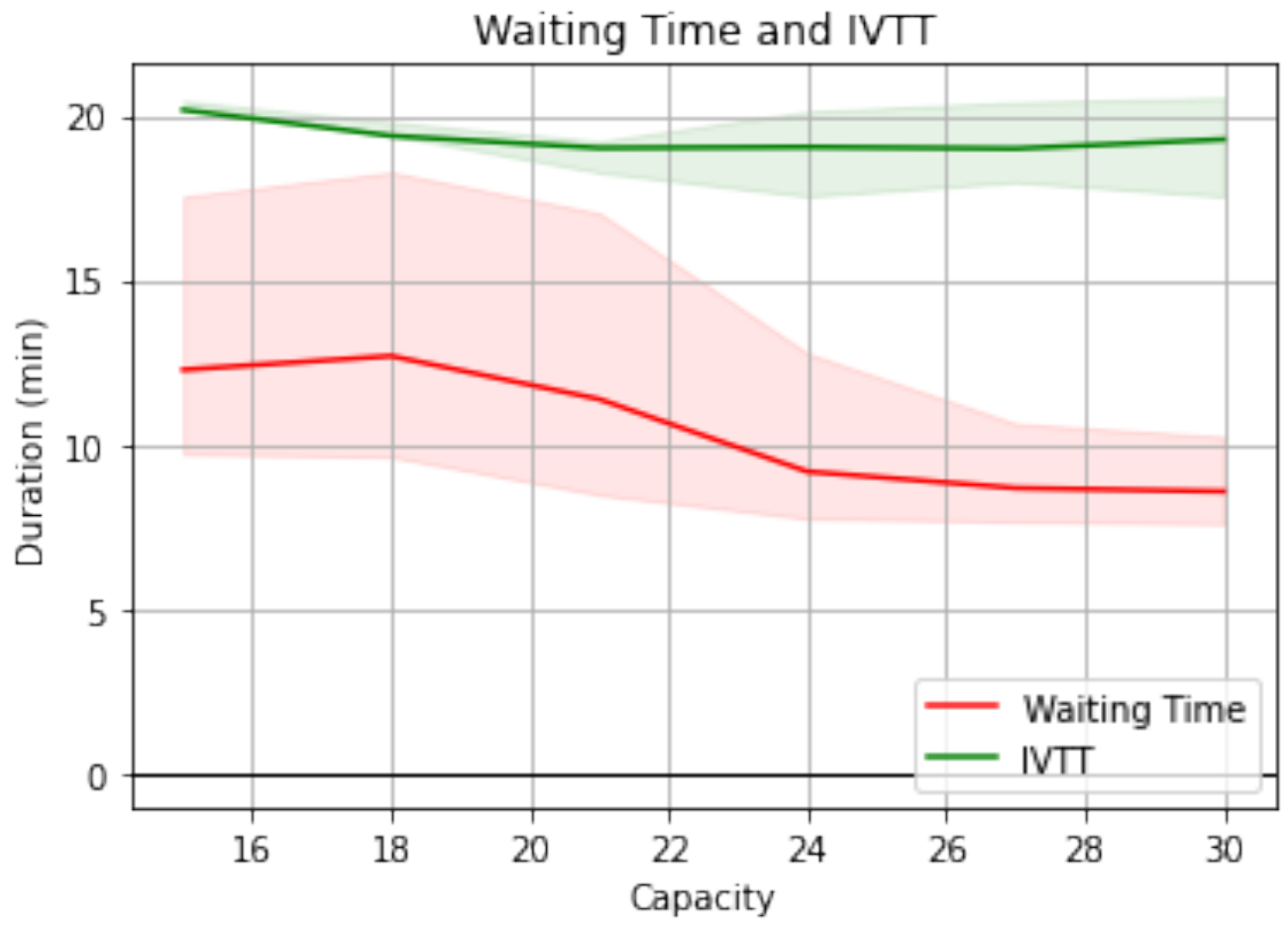

**Figure 26 - Mean Waiting Time and IVTT at Different Capacities**

*Sensitivity Analysis on Demand*

The difference in generalized costs shows a convex shape against passenger flow in **Figure 27**. The relative advantage of AMSoD increases when the number of passengers increases from 40 to 60, flattens till 80, and drops afterward. **Figure 28** shows strictly increasing waiting time and IVTT. This agrees with the results in the previous section, where Eq. (35) becomes $\lambda_k \leq 120$, *i.e.,* the semi-on-demand route is advantageous for hourly demand lower than 120.

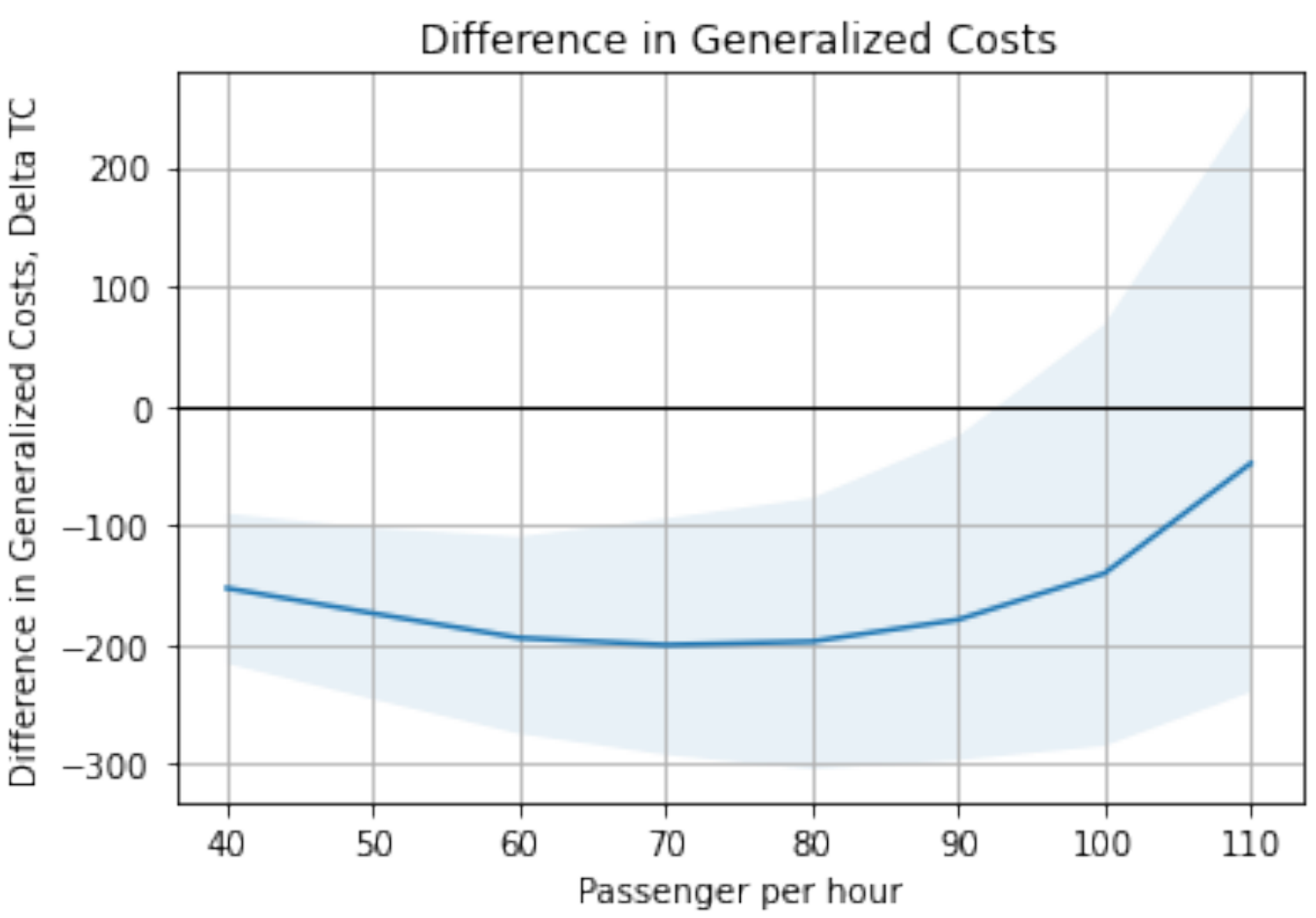

**Figure 27 - Difference in Generalized Costs at Different Hourly Demand**

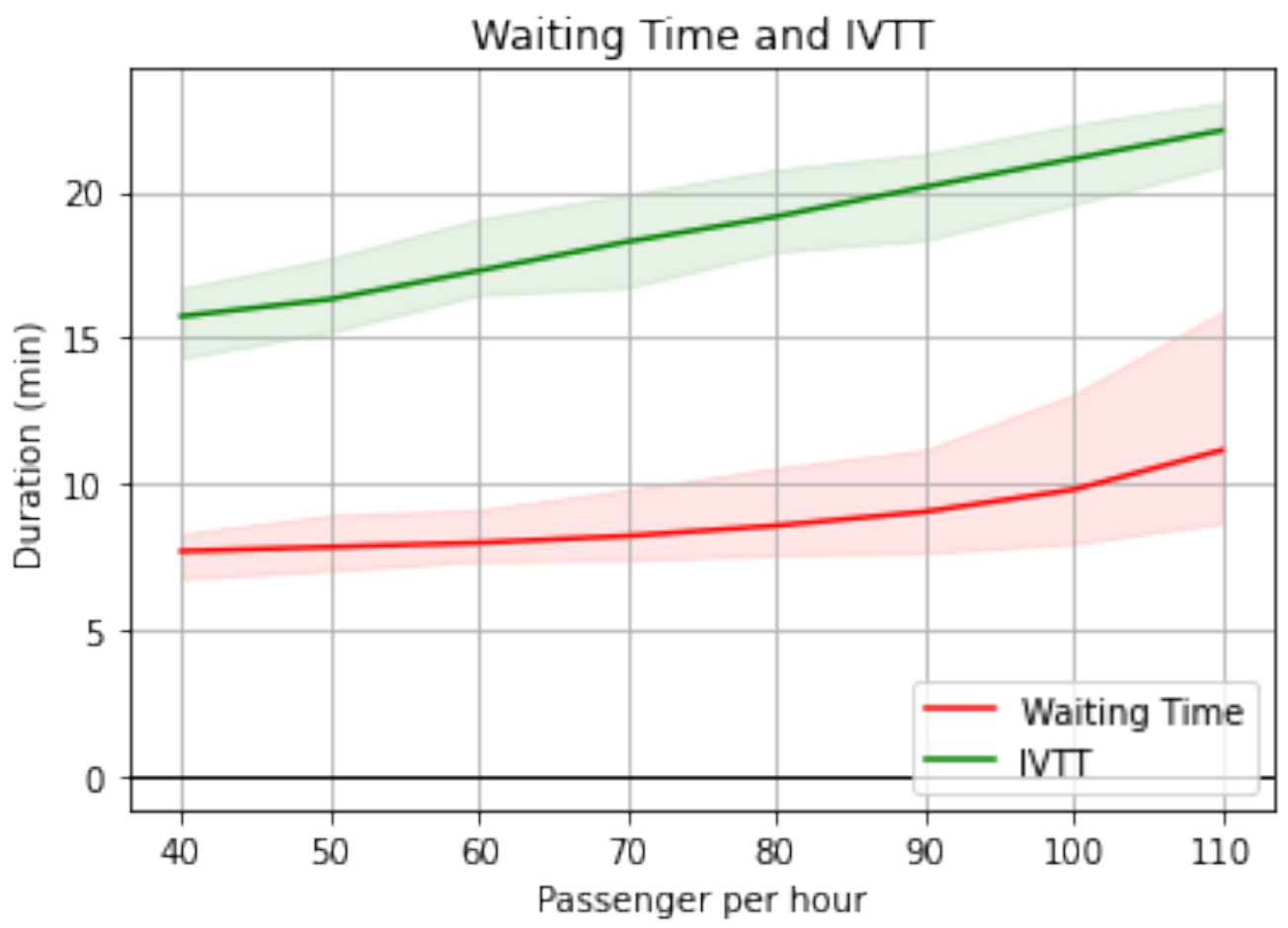

**Figure 28 - Mean Waiting Time and IVTT at Different Hourly Demand**

**CTA Bus #84**

We now look into an existing bus route with lower demand, the main portion of CTA Bus #84 along West Peterson Avenue eastbound connecting suburb to Red Line Bryn Mawr station (**Figure 21**). A similar approach of data collection and analysis is adopted. Periods of 9 a.m. – 1 p.m. and 5:55 p.m. – 9:30 p.m. are served with headway not smaller than 20min, indicating a perceived passenger flow of around 50/hour.

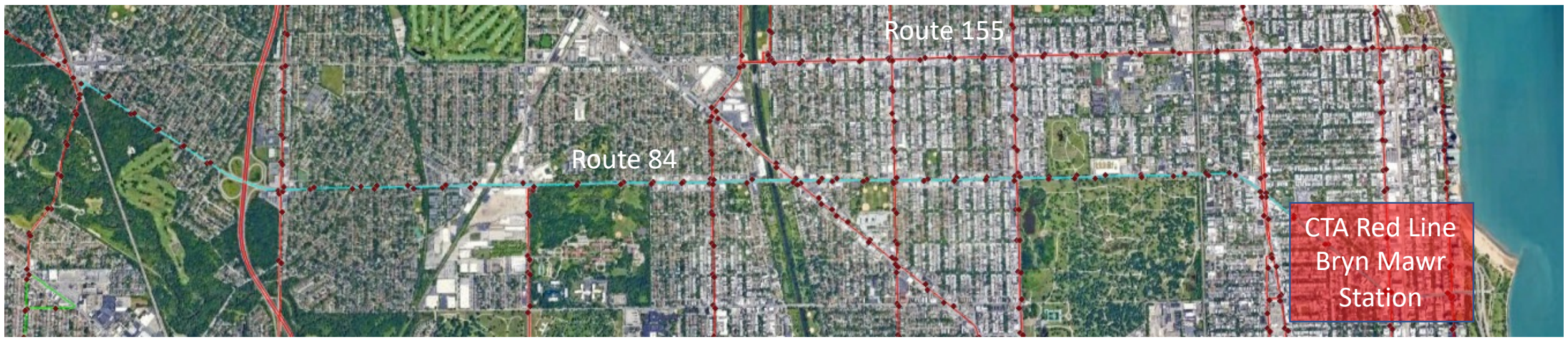

**Figure 29 - Map of CTA Bus #84**

A similar grid model is constructed (**Figure 22**). By Eq. (34) and $MD(Y) = 0.8\left(\frac{2}{3}\right) = 0.53$, $SI = 0.91 < 1$, indicating potential improvement.

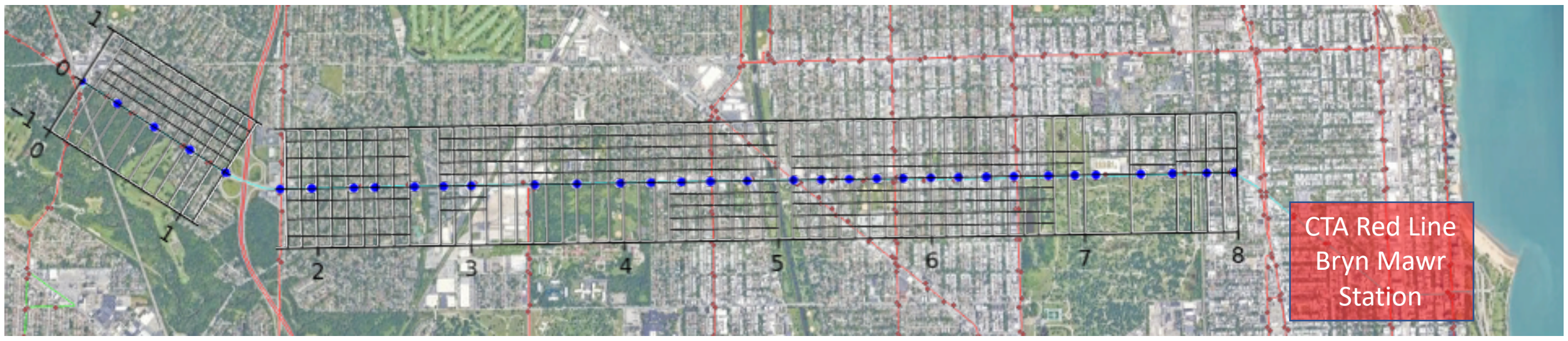

**Figure 30 - Grid model for CTA Bus #84**

*Modeling Result*

The average waiting time of semi-on-demand route is about 1min higher than that of the fixed route, similar to previous results despite the high y-directional dispersion for some parts of the route. This however causes a significant increase in average IVTT of 6.5min. The considerable saving in access cost balances all other costs, making AMSoD service favorable (**Figure 31**) with an equivalent $1.9 value for each passenger.

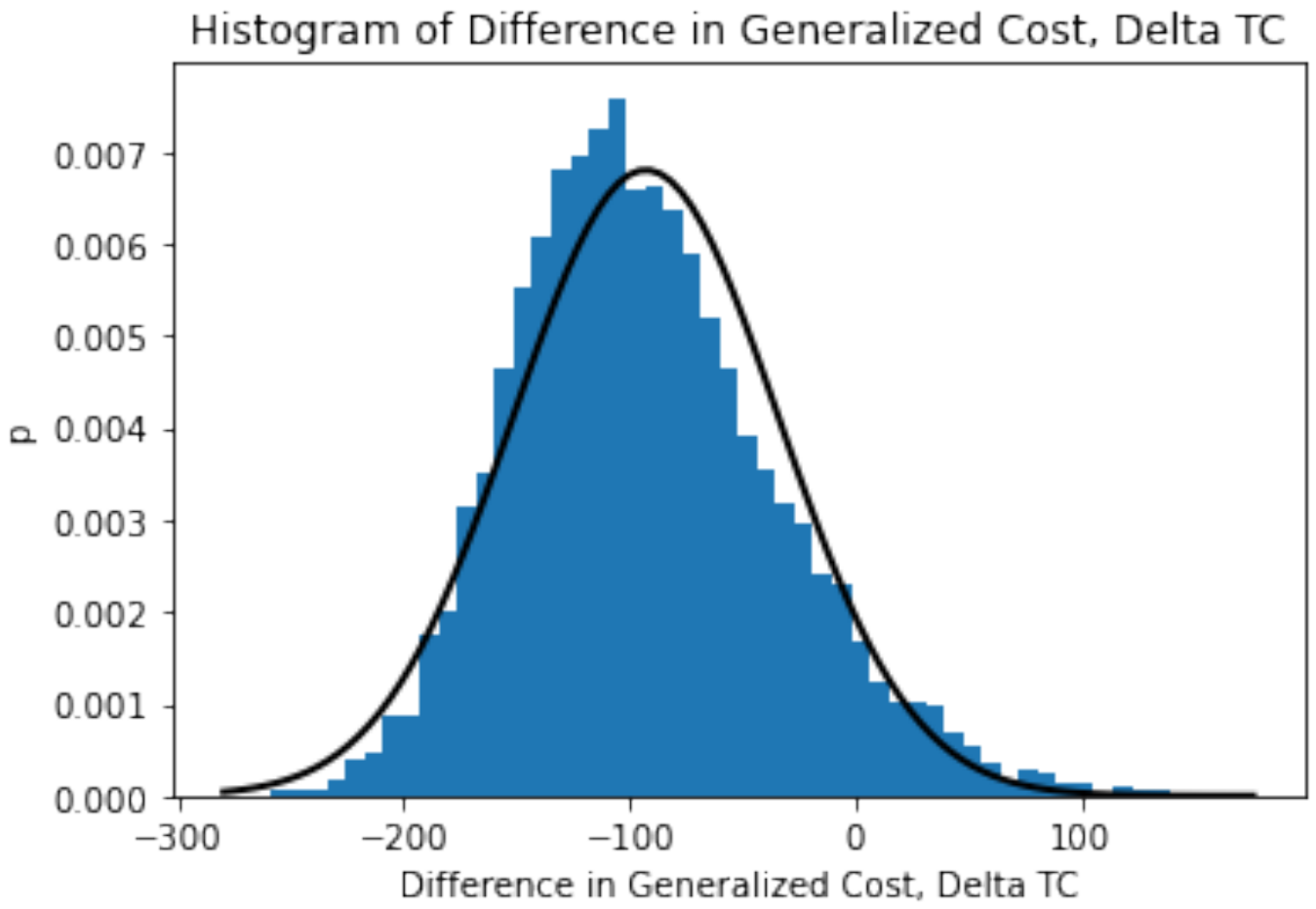

**Figure 31 - Histogram of Sum of Difference in Generalized Cost, $\sum \Delta TC$**

*Sensitivity Analysis on Bus Capacity*

While this case involves lower demand, the results shown in **Figure 32** and **Figure 33** are similar to previous models that minibuses with capacities higher than the threshold (demand) reduce waiting time.

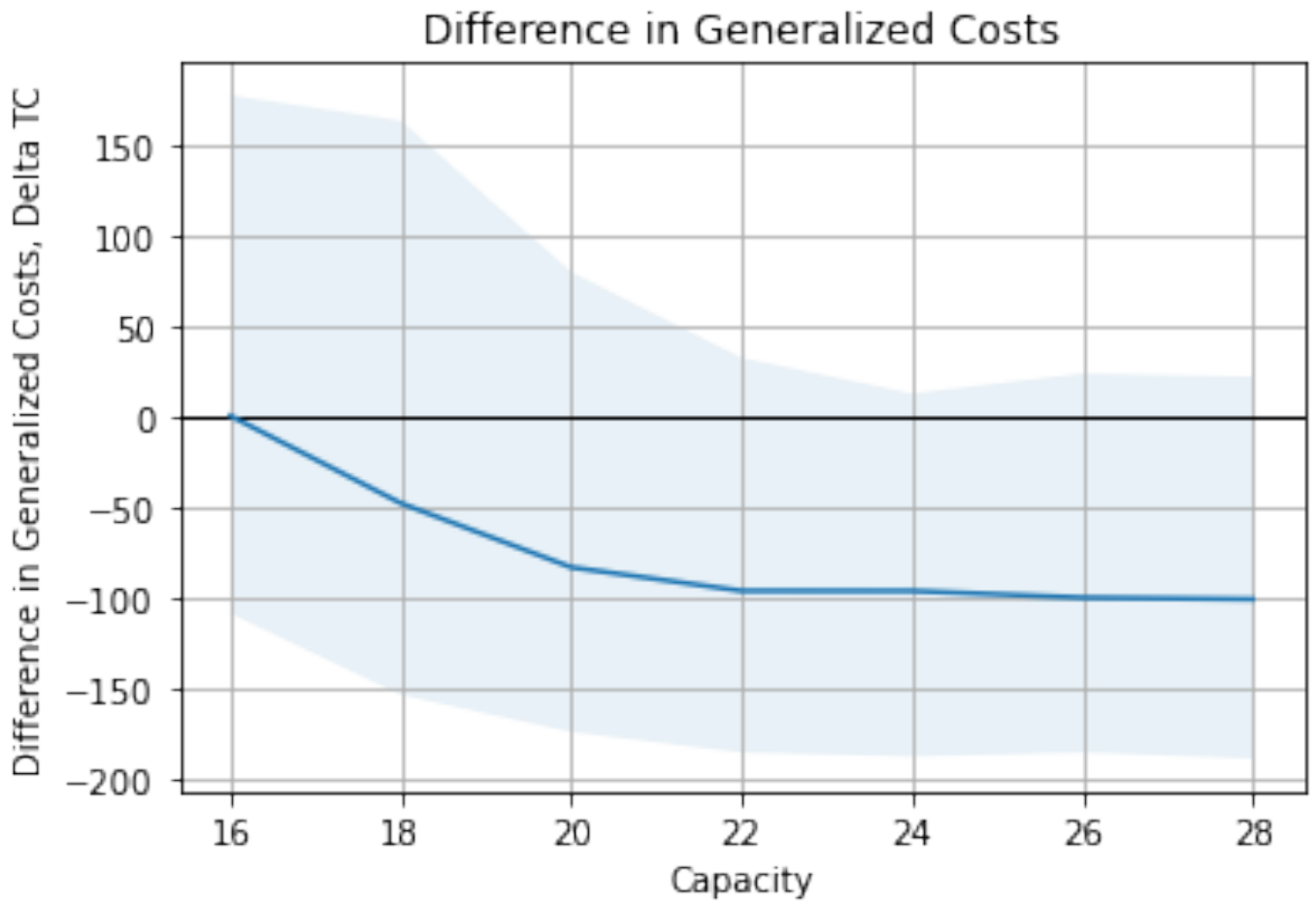

**Figure 32 - Difference in Generalized Costs at Different Capacities**

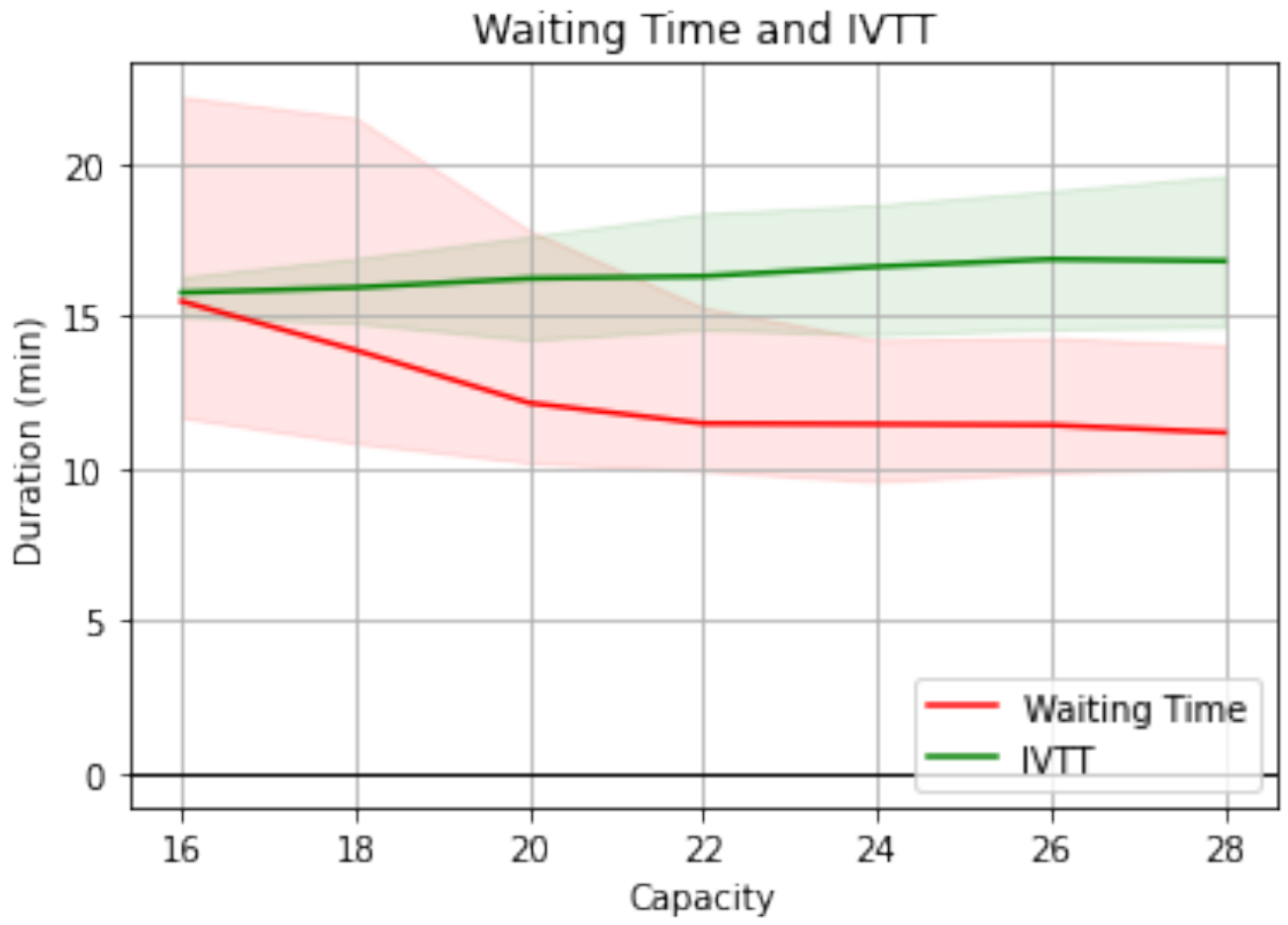

**Figure 33 - Mean Waiting Time and IVTT at Different Capacities**

*Sensitivity Analysis on Demand*

**Figure 34** shows that the demand of 50/hour in this case is close to optimum with similar generalized costs until 70/hour following the increase in IVTT in **Figure 35.** $\lambda_k \leq 65$ from Eq. (35) shares a similar result.

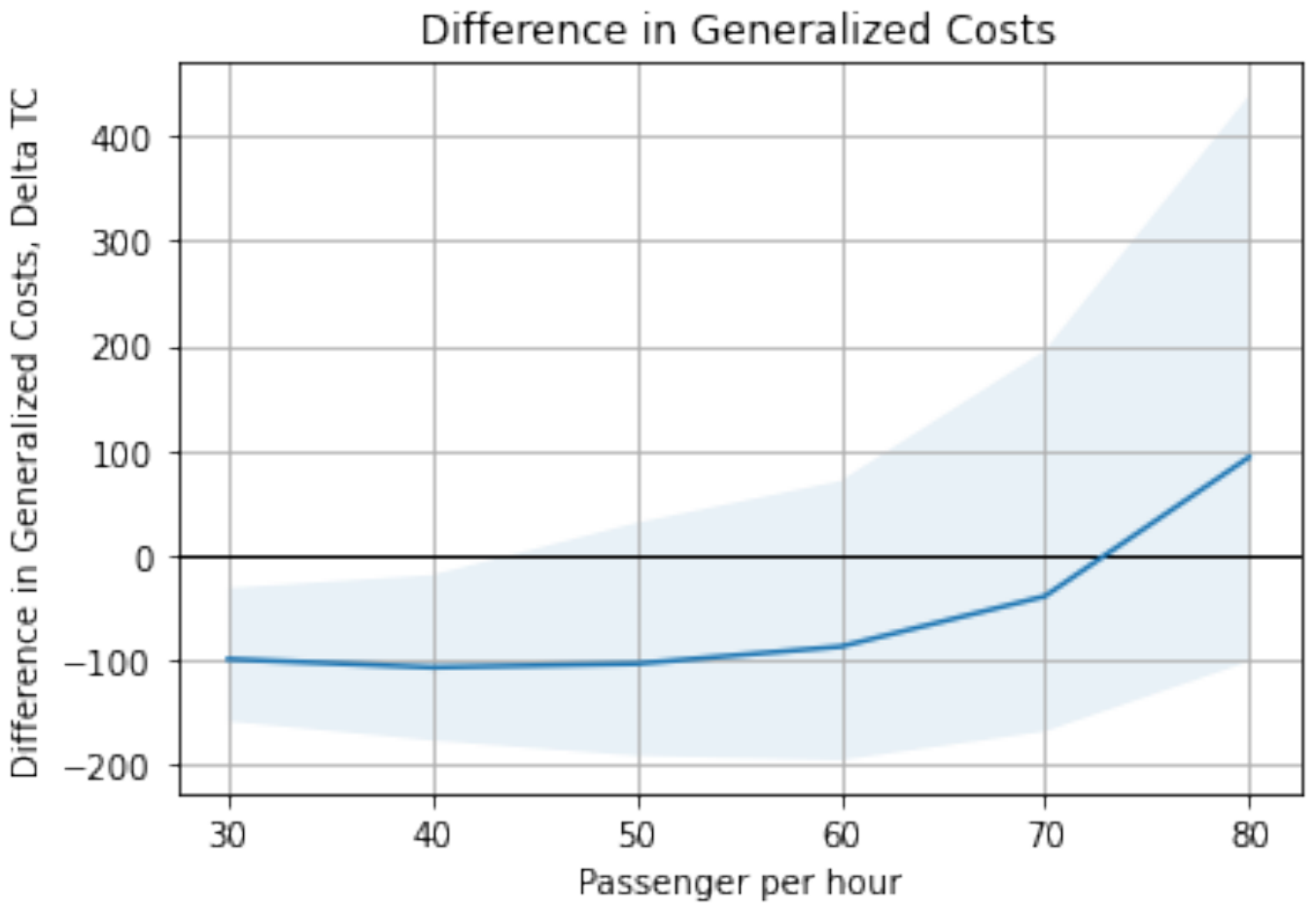

**Figure 34 - Difference in Generalized Costs at Different Hourly Demand**

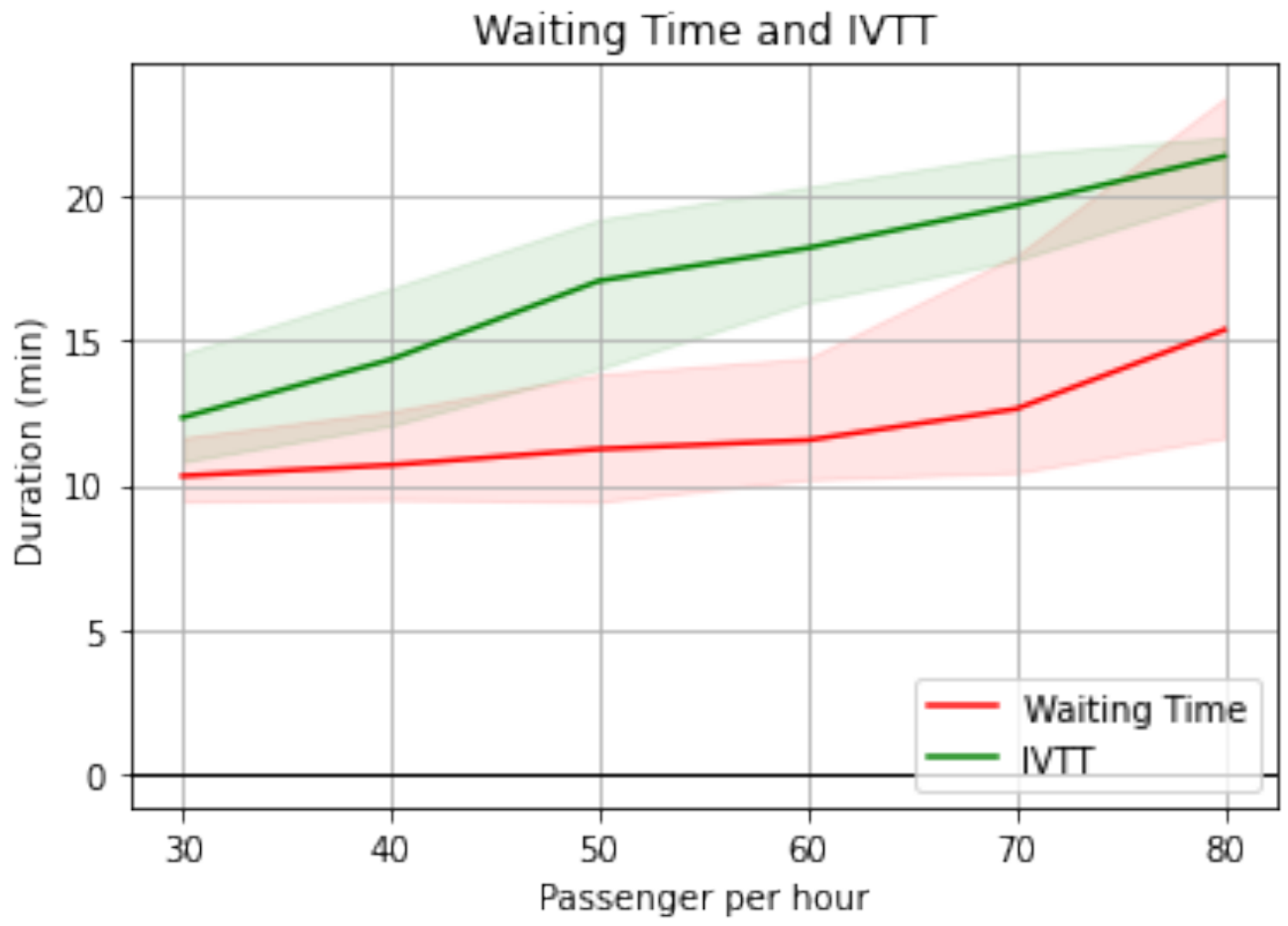

**Figure 35 - Mean Waiting Time and IVTT at Different Hourly Demand**

## CONCLUSION

This paper introduces the AMSoD service which combined the advantages of economy of scale in traditional buses and flexibility for passengers in DRT. Its costs in a grid system are derived with the options of zonal express. The benefits in directional routes from low-density areas (*e.g.,* suburb-CBD) over fixed routes in terms of generalized costs are demonstrated by mathematical formulation with parallel routes to cover more scenarios, implying higher attractiveness to passengers. Conversion criteria are proposed to identify the combinations of relatively low demand density and dispersion (MD) in which the proposed service can reduce generalized costs and offer higher attractiveness to existing passengers. Rule-based simulation with case study in Chicago supports the findings. Sensitivity analysis on capacity and demand shows that minibuses in capacities of 15-30 provide the best cost tradeoff with waiting time minimized. This aligns with its positioning with DRT and traditional fixed routes.

As for the second research question on how the AMSoD service can be expanded to cover more scenarios, zonal express takes advantage of higher hourly demand to separate the area into two or more routes. Expressions of the optimal number of zones are derived. Furthermore, parallel routes provide benefits in corridors with more widely distributed demand (*i.e.* larger MD) by reducing the additional y-directional distance traveled. The selection indicators are expanded to cater for the case of two parallel routes.

Although the developed model does not explicitly require autonomous functionality, we envision AV can support the service feasibility, addressing safety and reliability concerns to manage real-time requests and make immediate routing changes.

Nevertheless, this study is bounded by limitations, most of which are conservative assumptions and favor the AMSoD service:

1. Real-world demands are more likely concentrated, particularly along main streets, instead of distributed uniformly as assumed conservatively. The additional y-directional travel should be lower. Furthermore, concentrated demand in the x-direction is also beneficial by reducing such extra distance.
2. Additional demand induced outside due to the reduced access distance would attract higher patronage.
3. The actual waiting cost may be lower, because the AMSoD service provides a more certain arrival time by managing routing and passengers can wait at trip origins.
4. Group travelers would significantly reduce request points and y-directional detours.
5. Operators may have a set of preferable roads.
6. The AMSoD trips can be further optimized real-time by shortest path and strategic headway.
7. The potential change in IVTT and its variance may require more or fewer minibuses depending on its portion of the whole journey.
8. The bus is limited not to travel backward more than one block for pick-up (Service Assumption A7), which provides a worse performance than arranging routes to balance marginal benefits (waiting time saved) and costs (extra IVTT on bus and extra waiting time of later passengers).
9. A considerable portion of AMSoD benefits originates from saving in access costs, which are estimated from the value of access time. While a lower bound of such estimation still favors the proposed service, the perceived benefits may be reduced in extreme cases.

Further research efforts on the deployment and acceptance of autonomous minibus will be beneficial. More sophisticated models can be developed to incorporate group travel, agent-based simulation, and higher-level optimization considering wider transit network portions to select multiple routes for upgrading. Besides, the optimal bus capacity depends largely on fixed and variable costs of autonomous buses, which needs further forecasting efforts. It is also interesting to look at the change in interaction led by cost transfer between travelers and operators. Ultimately, this paper focuses on grid networks, which are common in American suburbs. More research on evaluating the benefits brought by the AMSoD service in other network types can widen its applicability.

## AUTHOR CONTRIBUTIONS

All authors contributed to study design, analysis and interpretation of results, and manuscript preparation. All authors reviewed the results and approved the submission of the manuscript.

## REFERENCES

1. Errico, F., T. G. Crainic, F. Malucelli, and M. Nonato. The Single-Line Design Problem for Demand-Adaptive Transit Systems: A Modeling Framework and Decomposition Approach for the Stationary-Demand Case. *Transportation Science*, Vol. 55, No. 6, 2021, pp. 1300–1321. https://doi.org/10.1287/trsc.2021.1062.

2. Errico, F., T. G. Crainic, F. Malucelli, and M. Nonato. A Survey on Planning Semi-Flexible Transit Systems: Methodological Issues and a Unifying Framework. *Transportation Research Part C: Emerging Technologies*, Vol. 36, 2013, pp. 324–338. https://doi.org/10.1016/j.trc.2013.08.010.
3. Frei, C. A. *User-Driven Demand Adaptive Transit: Enhancing the User Experience through Flexible Transit for Low-Density Communities*. Ph.D. Northwestern University, United States -- Illinois, 2015.
4. Cortés, C. E., and R. Jayakrishnan. Design and Operational Concepts of High-Coverage Point-to-Point Transit System. *Transportation Research Record*, Vol. 1783, No. 1, 2002, pp. 178–187. https://doi.org/10.3141/1783-22.
5. Nourbakhsh, S. M., and Y. Ouyang. A Structured Flexible Transit System for Low Demand Areas. *Transportation Research Part B: Methodological*, Vol. 46, No. 1, 2012, pp. 204–216. https://doi.org/10.1016/j.trb.2011.07.014.
6. Daganzo, C. F. Structure of Competitive Transit Networks. *Transportation Research Part B: Methodological*, Vol. 44, No. 4, 2010, pp. 434–446. https://doi.org/10.1016/j.trb.2009.11.001.
7. Sayarshad, H. R., and H. O. Gao. Optimizing Dynamic Switching between Fixed and Flexible Transit Services with an Idle-Vehicle Relocation Strategy and Reductions in Emissions. *Transportation Research Part A: Policy and Practice*, Vol. 135, 2020, pp. 198–214. https://doi.org/10.1016/j.tra.2020.03.006.
8. Chen, P. W., and Y. M. Nie. Analysis of an Idealized System of Demand Adaptive Paired-Line Hybrid Transit. *Transportation Research Part B: Methodological*, Vol. 102, 2017, pp. 38–54. https://doi.org/10.1016/j.trb.2017.05.004.
9. Calabrò, G., A. Araldo, S. Oh, R. Seshadri, G. Inturri, and M. Ben-Akiva. Integrating Fixed and Demand-Responsive Transportation for Flexible Transit Network Design. Presented at the TRB 2021: 100th Annual Meeting of the Transportation Research Board, 2021.
10. Narayanan, S., E. Chaniotakis, and C. Antoniou. Shared Autonomous Vehicle Services: A Comprehensive Review. *Transportation Research Part C: Emerging Technologies*, Vol. 111, 2020, pp. 255–293. https://doi.org/10.1016/j.trc.2019.12.008.
11. Alonso-Mora, J., S. Samaranayake, A. Wallar, E. Frazzoli, and D. Rus. On-Demand High-Capacity Ride-Sharing via Dynamic Trip-Vehicle Assignment. *Proceedings of the National Academy of Sciences*, Vol. 114, No. 3, 2017, pp. 462–467. https://doi.org/10.1073/pnas.1611675114.
12. Pinto, H. K. R. F., M. F. Hyland, H. S. Mahmassani, and I. Ö. Verbas. Joint Design of Multimodal Transit Networks and Shared Autonomous Mobility Fleets. *Transportation Research Part C: Emerging Technologies*, Vol. 113, 2020, pp. 2–20. https://doi.org/10.1016/j.trc.2019.06.010.
13. Winter, K., O. Cats, G. Correia, and B. van Arem. Performance Analysis and Fleet Requirements of Automated Demand-Responsive Transport Systems as an Urban Public Transport Service. *International Journal of Transportation Science and Technology*, Vol. 7, No. 2, 2018, pp. 151–167. https://doi.org/10.1016/j.ijtst.2018.04.004.
14. Dai, Z., X. C. Liu, X. Chen, and X. Ma. Joint Optimization of Scheduling and Capacity for Mixed Traffic with Autonomous and Human-Driven Buses: A Dynamic Programming

Approach. *Transportation Research Part C: Emerging Technologies*, Vol. 114, 2020, pp. 598–619. https://doi.org/10.1016/j.trc.2020.03.001.

15. Cao, Z., and A. (Avi) Ceder. Autonomous Shuttle Bus Service Timetabling and Vehicle Scheduling Using Skip-Stop Tactic. *Transportation Research Part C: Emerging Technologies*, Vol. 102, 2019, pp. 370–395. https://doi.org/10.1016/j.trc.2019.03.018.
16. Zhai, Z., Y. Yang, Y. Shen, Y. Ji, and Y. Du. Assessing the Impacts of Autonomous Bus-on-Demand Based on Agent-Based Simulation: A Case Study of Fuyang, Zhejiang, China. *Journal of Advanced Transportation*. Volume 2020, e7981791. https://www.hindawi.com/journals/jat/2020/7981791/. Accessed Jan. 15, 2021.
17. Salonen, A. O., and N. Haavisto. Towards Autonomous Transportation. Passengers' Experiences, Perceptions and Feelings in a Driverless Shuttle Bus in Finland. *Sustainability*, Vol. 11, No. 3, 2019, p. 588. https://doi.org/10.3390/su11030588.
18. Basu, R., A. Araldo, A. P. Akkinepally, B. H. Nahmias Biran, K. Basak, R. Seshadri, N. Deshmukh, N. Kumar, C. L. Azevedo, and M. Ben-Akiva. Automated Mobility-on-Demand vs. Mass Transit: A Multi-Modal Activity-Driven Agent-Based Simulation Approach. *Transportation Research Record*, Vol. 2672, No. 8, 2018, pp. 608–618. https://doi.org/10.1177/0361198118758630.
19. Bösch, P. M., F. Becker, H. Becker, and K. W. Axhausen. Cost-Based Analysis of Autonomous Mobility Services. *Transport Policy*, Vol. 64, 2018, pp. 76–91. https://doi.org/10.1016/j.tranpol.2017.09.005.
20. Pigeon, C., A. Alauzet, and L. Paire-Ficout. Factors of Acceptability, Acceptance and Usage for Non-Rail Autonomous Public Transport Vehicles: A Systematic Literature Review. *Transportation Research Part F: Traffic Psychology and Behaviour*, Vol. 81, 2021, pp. 251–270. https://doi.org/10.1016/j.trf.2021.06.008.
21. Bernhard, C., D. Oberfeld, C. Hoffmann, D. Weismüller, and H. Hecht. User Acceptance of Automated Public Transport: Valence of an Autonomous Minibus Experience. *Transportation Research Part F: Traffic Psychology and Behaviour*, Vol. 70, 2020, pp. 109–123. https://doi.org/10.1016/j.trf.2020.02.008.
22. Gschwender, A., S. Jara-Díaz, and C. Bravo. Feeder-Trunk or Direct Lines? Economies of Density, Transfer Costs and Transit Structure in an Urban Context. *Transportation Research Part A: Policy and Practice*, Vol. 88, 2016, pp. 209–222. https://doi.org/10.1016/j.tra.2016.03.001.
23. Zhang, J., D. Z. W. Wang, and M. Meng. Which Service Is Better on a Linear Travel Corridor: Park & Ride or on-Demand Public Bus? *Transportation Research Part A: Policy and Practice*, Vol. 118, 2018, pp. 803–818. https://doi.org/10.1016/j.tra.2018.10.003.
24. Edwards, D., and K. Watkins. Comparing Fixed-Route and Demand-Responsive Feeder Transit Systems in Real-World Settings. *Transportation Research Record: Journal of the Transportation Research Board*, Vol. 2352, 2013, pp. 128–135. https://doi.org/10.3141/2352-15.
25. Tong, L. (Carol), L. Zhou, J. Liu, and X. Zhou. Customized Bus Service Design for Jointly Optimizing Passenger-to-Vehicle Assignment and Vehicle Routing. *Transportation Research Part C: Emerging Technologies*, Vol. 85, 2017, pp. 451–475. https://doi.org/10.1016/j.trc.2017.09.022.
26. Newell, G. F. Some Issues Relating to the Optimal Design of Bus Routes. *Transportation Science*, Vol. 13, No. 1, 1979, pp. 20–35. https://doi.org/10.1287/trsc.13.1.20.

27. Wardman, M. Public Transport Values of Time. *Transport Policy*, Vol. 11, No. 4, 2004, pp. 363–377. https://doi.org/10.1016/j.tranpol.2004.05.001.
28. U.S. Census Bureau. Median Household Income in the United States. *FRED, Federal Reserve Bank of St. Louis*. https://fred.stlouisfed.org/series/MEHOINUSA646N. Accessed Feb. 28, 2021.
29. US Department of Transportation. The Value of Travel Time Savings: Departmental Guidance for Conducting Economic Evaluations Revision 2 (2016 Update). https://www.transportation.gov/office-policy/transportation-policy/revised-departmental-guidance-valuation-travel-time-economic. Accessed Feb. 27, 2021.
30. National Research Council (U.S.), Ed. *Bus Route and Schedule Planning Guidelines*. Transportation Research Board, National Research Council, Washington, D.C, 1980.
31. Chicago Transit Authority. CTA - Ridership - Avg. Weekday Bus Stop Boardings in October 2012. https://data.cityofchicago.org/Transportation/CTA-Ridership-Avg-Weekday-Bus-Stop-Boardings-in-Oc/mq3i-nnqe. Accessed Jan. 31, 2021.
32. Chicago Transit Authority. Service Standards. https://www.transitchicago.com/assets/1/6/servicestandards129737.pdf. Accessed Mar. 14, 2021.
33. TerraMetrics and NOAA. Google Earth Pro 7.3.3.7786. https://www.google.com/earth/index.html. Accessed Mar. 15, 2021.
34. Chicago Transit Authority. CTA - Bus Routes - Kml. https://data.cityofchicago.org/Transportation/CTA-Bus-Routes-kml/rytz-fq6y. Accessed Mar. 15, 2021.
35. Chicago Transit Authority. CTA - Bus Stops - Kml. https://data.cityofchicago.org/Transportation/CTA-Bus-Stops-kml/84eu-buny. Accessed Mar. 15, 2021.